\documentclass{article}

\usepackage[utf8]{inputenc}   
\usepackage[english]{babel}      
\usepackage{amsmath, amssymb, amsthm, mathrsfs}     
\usepackage[a4paper, margin=3cm]{geometry} 
\usepackage{graphicx}         
\usepackage{hyperref}         
\usepackage{tikz}
\allowdisplaybreaks
\usepackage{float} 
\usepackage{graphicx, pgfplots, xcolor}

\definecolor{granate}{RGB}{128, 0, 64} 

\usepackage[T1]{fontenc}
\usepackage{subcaption}
\usepackage{booktabs}

\newtheorem{theorem}{Theorem}

\newtheorem{corollary}[theorem]{Corollary}

\newtheorem{proposition}[theorem]{Proposition}

\begin{document}

\title{\textbf{Robust Estimation in Step-Stress Experiments under Exponential Lifetime Distributions.
		}}
\author{Mar\'{i}a Jaenada$^{(1)}, $ Juan Manuel Mill\'{a}n$^{(2)}$ and Leandro Pardo$^{(2)}$ \\
{\small $^{(1)}$ Department of Statistics, O.R. and N.A., UNED, Madrid, Spain}\\
{\small $^{(2)}$ Department of Statistics and O.R., Complutense University of Madrid, Spain}
}
\date{}
\maketitle

\begin{abstract}
	Many modern products exhibit high reliability, often resulting in long times to failure. Consequently, conducting experiments under normal operating conditions may require an impractically long duration to obtain sufficient failure data for reliable statistical inference. As an alternative, accelerated life tests (ALTs) are employed to induce earlier failures and thereby reduce testing time. In step-stress experiments a stress factor that accelerates product degradation is identified and systematically increased to provoke early failures. The stress level is increased at predetermined time points and maintained constant between these intervals. Failure data observed under increased levels of stress is statistically analyzed, and results are then extrapolate to normal operating conditions.
	
	Classical estimation methods such analysis rely on the maximum likelihood estimator (MLE) which is know to be very efficient, but lack robustness in the presence of outlying data. 
	In this work, Minimum Density Power Divergence Estimators (MDPDEs) are proposed as a robust alternative, demonstrating an appealing compromise between efficiency and robustness. The MDPDE
	based on mixed distributions is developed, and its theoretical properties, including the expression for the asymptotic distribution of the model parameters, are derived under exponential lifetime assumptions. The good performance of the proposed method is evaluated through simulation studies, and its applicability is demonstrated using real data.
\end{abstract}

\bigskip \underline{\textbf{AMS 2001 Subject Classification}}\textbf{: }


\section{ Introduction\label{sec:intro}}

Reliability analyses aim to infer the lifetime behavior of a device or system. That is, the time duration until it ceases to function under normal operating conditions. Many modern products and systems are designed with long mean lifetimes to failure (MTTF), often extending over several years. Under these circumstances, obtaining sufficient lifetime observations for statistical inference under operating conditions becomes challenging, making it difficult to draw conclusions about the life behaviour of the device.

To address this challenge, accelerated life tests (ALTs) are applied, by subjecting the test devices or systems to harshest than normal physical conditions that wear down the product and thus reduce it time to failure. 
For example, experimenter may increase environmental conditions such as temperature, voltage, pressure, load, humidity, or some combination of them to induce the wear and tear of the device.
ALT models need a model describing the relationship between lifetime behavior and stress load, allowing the extrapolation of results obtained under accelerated conditions to normal operating conditions.
 Some interesting works in ALTs can be seen in Nelson (2005a, b, 2009), Limon et al (2017), Chen et al (2018), Escobar, and Meeker (2006), Pinto-Santos et al (2021). 
There exist mainly three types of accelerated life testing designs used in practice, depending on how the stress is applied; Constant stress ALT (CSALTs), Step-stress ALTs (SSALTs) and progressive stress ALTs (PSALTs). 
Each stress design has its own advantages and disadvantages, for example, Nelson (1990) demonstrates that SSALTs may provide more accurate estimates that CSALTs under similar experimental conditions.
The choice of the experimental design will also depend on the nature of the product and the specific experimental requirements and limitations.
In this paper we will consider SSALTs.

\begin{figure}[h!]
\centering
\begin{tikzpicture}[scale=0.7] 
  \draw[->] (0, 0) -- (6, 0) node[below] {\small Time}; 
  \draw[->] (0, 0) -- (0, 3.5) node[left] {\small Stress Level}; 

  \draw[thick, gray] (0, 1) -- (2, 1) node[pos=0.5, below] {$x_1$};
  \draw[thick, purple] (2, 2) -- (5, 2) node[pos=0.5, below] {$x_2$};

  \fill[blue] (0, 1) circle(2pt);
  \fill[blue] (2, 1) circle(2pt);
  \fill[blue] (2, 2) circle(2pt);
  \fill[blue] (5, 2) circle(2pt);

  \node at (2, -0.3) {$\tau_1$};
  \node at (5, -0.3) {$\tau_2$};

  \draw[dashed] (2, 0) -- (2, 2);
  \draw[dashed] (5, 0) -- (5, 2);

\end{tikzpicture}
  \caption{SSALT design under Type I censoring.}
  \label{fig:step-stress}
\end{figure}
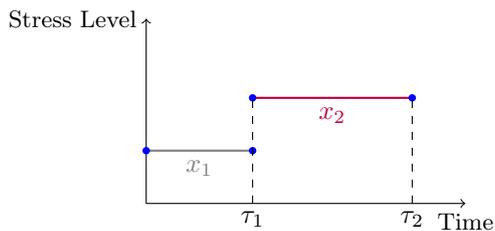

In SSALTs, the experiment is structured as follows: the total experimental duration is divided into $2$ stress intervals, defined by the time points \( \tau_0=0, \tau_1\) and \(\tau_2\) . During each interval, a different stress level, denoted by \(x_1\) and \(x_2\), is applied to the test units.
A total of \( N \) products are tested under varying stress levels, and the failure times of the tested devices are recorded, \( t_j, j =1, ..., N \).
Moreover, the stress level is assumed to be gradually increasing, in line with the premise that the product undergoes increasing wear to induce failure.
The experiment concludes at a fixed time \( \tau_2 \), when some devices may not have failed within the total duration of the test and so surviving units are right-censored. This set-up corresponds to Type-I censoring. An scheme of the experimental design is shown in figure \ref{fig:step-stress}.

\par

Besides, we denote by \( n_i \) the number of failed devices under $x_i$ stress level, $i=1, 2$. By the end of the experiment, the total number of surviving units will be:
\(
N - \sum_{i=1}^{2} n_i.
\)
Then, the observed failure times are ordered as follows:
\begin{equation}
\tau_0 = 0 < t_{1:N} \dots <t_{n_1:N} < \tau_1 < t_{n_1+1:N} \dots <t_{n_1+n_2:N}< \tau_2.
\label{eq:order}
\end{equation}

We will assume exponential distributions for the devices lifetimes. It has only one model parameter, $\lambda$, the rate of distribution.
Given a random variable T and the rate parameter $\lambda$, the probability distribution function of the exponential distribution is:
\[
F_T(t|\lambda)=1-\exp\left(-\frac{t}{\lambda}\right), \quad t>0,
\]
and the probability density function (p.d.f.) is:
\[
f_T(t|\lambda)=\frac{1}{\lambda}\exp\left(-\frac{t}{\lambda}\right), \quad t>0, \quad \lambda>0.
\]
\\
%
%
%
%
%

SSALTs require a model relating the lifetime distribution at a current stress level to the proceeding levels. There are three fundamental models for the effect of increased stress levels on the lifetime distribution of a device: The tampered random variable (TRV) model, proposed by DeGroot and Goel (1979), the cumulative exposure (CE) model of Sedyakin (1966) and Nelson (1980), and the tampered failure rate model (TRM), proposed by Bhattacharyya and Soejoeti (1989). 
The CE assumes that the wear accumulated by a product under a given stress level carries over to the next level as a shift in the reliability of the devices, without retaining memory of how the wear occurred. This assumption is intuitive and has been widely adopted in the literature. Moreover, under exponential distributions, all the proposed models are equivalent (see Rao (1992)). Thus, we develop the theory under a CE approach here, but all results would be valid under TRV and TRM.

Under the CE model, the cumulative distribution function (c.d.f.) of the failure time is a continuous, piecewise-defined function, where failure times follow exponential distributions under each stress level. Wear and tear accumulated at preceding stresses is transferred to subsequent stress levels by shifting the distribution at each level. Thus, the c.d.f. of the devices lifetimes under simple SSALT, $G_T$ is given by 
\begin{equation}
	G_T(t) = \begin{cases}
		G_1(t) & 0 < t < \tau_1\\
		G_2(t+h) & \tau_1 \leq t < \tau_2
	\end{cases}
\end{equation}
where the shifting time $h$ ensures the continuity of the c.d.f.,
\begin{equation*}
G_{1}(\tau_1) = G_{2}(\tau_2 + h).
\end{equation*}
The cumulative distribution function can be extended to infinity by simply setting $\tau_2=\inf.$

Under exponential lifetimes, the c.d.f. under constant stress levels, \( G_1(t) \) and \( G_2(t) \), are defined by scale parameters \( \lambda_1 \) and \( \lambda_2 \), respectively. Therefore, using the properties of the scale parameters, the shifting time \( h \) can be expressed in closed form as follows:
\begin{equation}
h = \frac{\lambda_2}{\lambda_1} \tau_1 - \tau_1. 
\end{equation}

Thus, the c.d.f. of the devices lifetimes is given by
\begin{equation}
G_T(t|\boldsymbol{\lambda}) =
\begin{cases}
 G_1(t|\lambda_1)=  1 - \exp\left(-\frac{t}{\lambda_1}\right) \quad &0 < t < \tau_1 \\
  G_2(t|\lambda_1, \lambda_2)=  1 - \exp\left(-\frac{t+h}{\lambda_2}\right) \quad &\tau_1 < t < \infty
\end{cases},
\end{equation}
with $\boldsymbol{\lambda} = (\lambda_1, \lambda_2) \in \mathbb{R}^+ \times \mathbb{R}^+$ and the corresponding p.d.f. is given by
\begin{equation}
g_T(t|\boldsymbol{\lambda}) =
\begin{cases}
 g_1(t|\lambda_1)= \frac{1}{\lambda_1} \exp\left(-\frac{t}{\lambda_1}\right) \quad &0 \leq t < \tau_1 \\
  g_2(t|\lambda_1, \lambda_2)= \frac{1}{\lambda_2} \exp\left(-\frac{t + h}{\lambda_2}\right) \ &\tau_1 \leq t < \infty
\end{cases}.
\end{equation}

Based on the assumption that increasing stress accelerates product wear, the failure time distribution under constant stress should depend on the stress level. Therefore, it is necessary to establish a relationship between the exponential distribution parameter and the applied stress. For such a purpose, a log-logistic relationship is commonly adopted,
\begin{equation}
\lambda_i = \exp(a_0 + a_1 x_i)\ \text{for } i=1, 2,
\label{eq:funcrelation}
\end{equation}
where $(a_0, a_1) \in \mathbb{R}\times \left(-\infty, 0\right)$ is the regression vector. We assume that $a_1$ is negative, as increased stress is expected to lead to a reduction in the mean lifetime.

Given the failure times \( t_{1:N}, \dots, t_{n_1:N}, t_{n_1+1:N}, \dots, t_{n_1+n_2:N} \), the log-likelihood of the SSALT model 
is given by
\begin{equation}
	\begin{aligned}
	\ell(a_0, a_1) =&\log\left(N!\right) -\log \left((N-n_1-n_2)!\right)\\
	& + \sum_{i=1}^{n_1} \log\left(g_1(t_i|\lambda_1)\right) \\
	& + \sum_{i=n_1+1}^{n_1+n_2} \log(\left(g_2(t_i + h|\lambda_2)\right) \\
	& +(N-n_1-n_2)\log(\left( 1 - G_2(\tau_2 + h|\lambda_2) \right),
	\end{aligned}
	\label{eq:likelihood}
\end{equation}
where \[ \lambda_i = \exp(a_0+a_1x_i).\]


Therefore, the MLE for the SSALT is defined as
\begin{equation}
\left(\hat{a}_0^{MLE}, \hat{a}_1^{MLE} \right) = \arg \max_{(a_0, a_1) \in A} \ell (a_0, a_1),
\end{equation}
with
 $A=\left\{ (a_0, a_1) \in \mathbb{R} \times \left(-\infty, 0\right)\right\}$. Importantly, note that the MLEs for \( \lambda_1 \) and \( \lambda_2 \) can only be obtained simultaneously if \( n_1 \geq 1 \) and \( n_2 \geq 1 \). Therefore, the existence of the MLE depends on having failures in both intervals. This condition will be assumed throughout the article. 

Under exponential distributions and assuming log-logistic relationship as defined in \eqref{eq:funcrelation}, the observed log-likelihood function for the SSALT model is given by
\begin{equation}
	\begin{aligned}
			\ell(a_0, a_1) = &
			\log\left(N!\right)-\log\left((N-n_1-n_2)!\right) -n_1\log\left(\lambda_1\right)-n_2\log\left(\lambda_2\right)\\
			& -\sum_{i=1}^{n_1}\frac{t_i}{\lambda_1} - \sum_{i=n_1+1}^{n_1+n_2}\frac{t_i+h}{\lambda_2} -(N-n_1-n_2)\frac{\tau_2+h}{\lambda_2}.
	\end{aligned}
\label{eq:log_likelihood}
\end{equation}

It is well known that the MLE is a BAN estimator (Best Asymptotically Normal), i.e., it is asymptotically efficient. Conversely, it is heavily influenced by outlying observations. 
To address this problem, several robust estimators have emerged in the statistical literature, mitigating the influence of outlier observations.
A well-known family of robust estimators is obtained by minimizing the density power divergence (DPD) between two appropriate distributions. 
This family of divergence measures was considered for the first time in Basu et al.(1998) and, from there, a wide literature has adopted the DPD approach, proven its robustness with small loss of efficiency. In particular, the MDPDE has been used in reliability theory, in general, and under a step-stress model with interval-censoring (see for example Balakrishnan et al 2022, 2023a, b, c, 2024). 
In these previous works the step-stress model was discretizated and the estimators of the unknown parameters are estimated by using the multinomial model associated. 
However, the discretization of the model results in a important loss of information, as the parameter estimates are obtained based on only on counts of failures, and not on the exact failure times observed. In this work we consider, for the first time in the literature, robust estimators based on the DPD using exact failure data.

In Section \ref{sec:MDPDE} the MLEs of the exponential parameters are defined by using  the Kullback-Leibler divergence between two mixed distributions. Later, the MDPDE is presented as a natural extension of the MLE, but relying on a tuning parameter that controls the trade-off between robustness and efficiency.
 The asymptotic distribution of the MDPDEs is obtained in Section \ref{sec:asymptotic} and asymptotic confidence intervals are later obtained in Section \ref{sec:CI}. In Section \ref{sec:simulation} a simulation study is carried out to evaluate the robustness of the MDPDE. A real dataset is analyzed in Section \ref{sec:example} to illustrate the applicability of the method. Finally, Section \ref{sec:concluding} presents some concluding remarks and future works lines.

\section{The minimum density power divergence estimator \label{sec:MDPDE}}

\bigskip

In this section, the MLE is obtained as minimum divergence estimator, by minimizing the Kullback-Leibler (KL) divergence. 
Before introducing the KL divergence, let us first justify the use of mixed distributions. Although devices under test are continuously monitored during the experiment,
once the experiment concludes, only the number of surviving devices is observed, lacking information on the specific time of failure of the surviving devices. Consequently, the observable 
distribution, in the following denoted by $F_T$, is a mixed distribution consisting of a continuous distribution between \( (0, \tau_2) \), and a discrete mass function at \( \tau_2 \), given by \( 1 - F_T(\tau_2) \). This probability corresponds to the probability of survival up to the end of the experiment.
That is, the observable c.d.f. under exponential lifetimes is given by
\begin{equation}
F_T(t|\boldsymbol{\lambda})=
\begin{cases}
0 & \quad t<0
\\
F_{1}(t|\lambda_1)=1-\exp\left(-\frac{t}{\lambda_1} \right) \quad& 0\leq t <\tau_1 
\\
F_{2}(t+h|\boldsymbol{\lambda})=1-\exp\left(-\frac{1}{\lambda_2}\left(t+h\right)\right) \quad &\tau_1 \leq t < \tau_2
\\
1 \quad & \tau_2 \leq t
\end{cases}.
\label{eq:funF}
\end{equation}
Where $\boldsymbol{\lambda} =\left(\lambda_1, \lambda_2 \right) $ and $\lambda_1, \lambda_2$ where defined in \eqref{eq:funcrelation}.
Similarly, the true underlying c.d.f. of the lifetime, truncated at the end of the experiment will also be given as a piecewise distribution of the form
\begin{equation}
	G_T(t)=
	\begin{cases}
		0 \quad & t<0
		\\
		G_1(t) \quad & 0 \leq t <\tau_1
		\\
		G_2(t) \quad & \tau_1 \leq t < \tau_2
		\\
		1 \quad & \tau_2 \leq t
	\end{cases},
\label{eq:funG}
\end{equation}
where $G_1(\cdot)$ and $G_2(\cdot)$ denote the c.d.f.s under constant stresses $x_1$ and $x_2$, respectively.
Note that, rather than assuming a specific model relating the c.d.f. at different stresses, we only consider that it may change when the stress level varies.
\\
Besides, for mathematical simplicity, in the following we denote,
\begin{align*}
	f_1(t|\lambda_1)=\frac{\partial F_1(t|\lambda_1)}{\partial t} \quad \quad g_1(t)=\frac{\partial G_1(t)}{\partial t} \quad & 0 < t <\tau_1 
	\\
	f_2(t+h|\boldsymbol{\lambda})=\frac{\partial F_2(t+h|\boldsymbol{\lambda})}{\partial t} \quad \quad g_2(t)=\frac{\partial G_2(t)}{\partial t} \quad & \tau_1 < t <\tau_2 .
\end{align*}
Where $\boldsymbol{\lambda} =\left(\lambda_1, \lambda_2 \right) $ is defined in \eqref{eq:funcrelation}.These functions are not proper density functions themselves, but are proportional to them up to a constant factor, as they represent the continuous component of the mixed distribution.
\\

From the above, we can define the Kullback-Leibler divergence between the mixed distributions $F_T(\cdot|\boldsymbol{\lambda})$ and $G_T(\cdot)$ as defined in \eqref{eq:funF} and \eqref{eq:funG} is given by:
\begin{equation}
	\begin{aligned}
		d_{KL}\left(G_T\left(\cdot\right), F_T\left(\cdot|\boldsymbol{\lambda}\right)\right)=&\int_0^{\tau_1} g_1(t) \log\left(\frac{g_1(t)}{f_1(t|\lambda_1)}\right) dt \int_{\tau_1}^{\tau_2}g_2(t)\log\left(\frac{g_2(t)}{f_2( t+h|\boldsymbol{\lambda})}\right)dt\\
		&+\left(1-G_2\left(\tau_2\right)\right)\log\left(\frac{1-G_2(\tau_2)}{1-F_2(\tau_2+h|\boldsymbol{\lambda})}\right).
	\end{aligned}
\label{eq:kl_def}
\end{equation}
The idea of defining the KL divergence for mixed distributions as a combination of its continuous and discrete components is not novel, as a similar approach has been proposed previously in the case of the Shannon entropy (see Nair et al (2007)).

After some algebra, expression \eqref{eq:kl_def} can be written as follows:
\begin{equation}
	\begin{aligned}
		d_{KL}\left(G_T\left(\cdot\right), F_T\left(\cdot|\boldsymbol{\lambda}\right)\right)=&-\int_0^{\tau_1} \log\left( f_1(t|\lambda_1)\right)dG_1(t)-\int_{\tau_1}^{\tau_2}\log\left(f_2(t+h|\boldsymbol{\lambda})\right)dG_2(t)
		\\
		&-\left(1-G_2\left(\tau_2\right)\right)\log\left(1-F_2(\tau_2+h|\boldsymbol{\lambda}\right)+K,
	\end{aligned}
\end{equation}
where K a constant that does not deppend on $\lambda$.

Our goal is to find the best estimate of $\boldsymbol{\lambda}$ (or $\boldsymbol{a} = (a_0, a_1)$ if the log-linear relationship holds), such that the assumed parametric form of the distribution fits the true distribution as closely as possible. Because the true distribution functions $G_1(\cdot)$ and $G_2(\cdot)$ are unknown and therefore, given a random sample from the experiment, $(t_{i:N}, ..., t_{n1+n2:N})$, we can approximate them by the corresponding empirical distribution functions associated. Then we have
\begin{equation}
	\begin{aligned}
		d_{KL}\left(G_T\left(\cdot\right), F_T\left(\cdot|\boldsymbol{\lambda}\right)\right)\approx&-\frac{n_1}{N}\sum_{i=1}^{n_1} f_1(t_{i:N}|\lambda_1)-\frac{n_2}{N}\sum_{i=n_1+1}^{n_1+n_2}\log\left(f_2(t_{i:N}+h|\boldsymbol{\lambda})\right)
		\\
		&-\frac{N-n_1-n_2}{N}\log\left(1-F_2(\tau_2+h|\boldsymbol{\lambda}\right)+K.
	\end{aligned}
\end{equation}
In the case of the exponential lifetimes, the explicit expression of the KL is as follows:

\begin{equation}
	 \begin{aligned}
		 d_{\text{KL}}\left(G_T(\cdot), F_T(\cdot|\boldsymbol{\lambda})\right)&= -\frac{n_1}{N} \log(\lambda_1) - \frac{n_1}{N}\frac{1}{\lambda_1} \sum_{i=1}^{n_1} t_i 
		 \\
		 &- \frac{n_2}{N}\log(\lambda_2)- \frac{n_2}{N}\frac{1}{\lambda_2}\sum_{i=n_1+1}^{n_{1}+n_2} ( t_i +h)
		\\
		 &- \frac{N - n_1 - n_2}{N} \left(\frac{1}{\lambda_2}(\tau_2+h)\right)+K
		 \\
		 &= K - \frac{1}{N} \ell(\lambda_1, \lambda_2)
	\end{aligned}
\end{equation}
where $\ell(\lambda_1, \lambda_2)$ is the log-likelihood of the model, as defined in \eqref{eq:log_likelihood}.
Therefore, the values of $(\lambda_1, \lambda_2)$ that minimizes the estimated Kullback-Leibler divergence is the same as the one that maximizes the log-likelihood function. Thus, assuming the log-linear relationship and reparametrizing accordingly, we can equivalently defined the MLE as the minimum KL estimator,
\begin{equation}
\left(\hat{a}_0^{MLE}, \hat{a}_1^{MLE}\right)= \arg \min_{a_0, a_1 \in \mathbb{R}^2}d_{\text{KL}} \left(G_T(\cdot)^*, F_T(\cdot|\boldsymbol{\lambda})\right).
\label{eq:kl_min}
\end{equation}
The DPD is a widely used divergence measure for developing robust methods of statistical inference. It provides a balance between efficiency and robustness, allowing for a controlled trade-off through a tuning parameter 
$\beta$. Introduced by Basu et al. (1998), the DPD has been extensively used thanks to its ability to produce estimators that maintain high efficiency while significantly improving robustness against outliers of statistical models.
\\

Following a similar approach to that used for the KL divergence for mixed distributions, we define the DPD between $G(\cdot)$ and $F(\cdot|\boldsymbol{\lambda})$ as a combination of the DPD between both continuos and discrete components as follows:
\begin{equation}
	\begin{aligned}
		d_{\beta}\left(G(\cdot), F(\cdot|\boldsymbol{\lambda})\right) &=\int_0^{\tau_1} \left(f_1(t|\lambda_1)^{\beta+1} - \left(1+\frac{1}{\beta} \right) f_1(t|\lambda_1)^{\beta} g_{1}(t)+\frac{1}{\beta}g_{1}(t)^{\beta+1} \right) dt
		\\
		&+\int_{\tau_1}^{\tau_2} \left(f_2(t+h|\boldsymbol{\lambda})^{\beta+1} - \left(1+\frac{1}{\beta} \right) f_2(t+h|\boldsymbol{\lambda})^{\beta} g_{2}(t)+\frac{1}{\beta}g_{2}(t)^{\beta+1} \right)dt
		\\
		&+\left(1-F_2(\tau_2+h|\boldsymbol{\lambda})\right)^{\beta+1}-\left(1+\frac{1}{\beta}\right)\left(1-F_2(\tau_2+h|\boldsymbol{\lambda})^{\beta}\right)\left(1-G_{2}(\tau_2)\right)
		\\
		&+\frac{1}{\beta}\left(1-G_{2}(\tau_2)\right)^{\beta+1}.
	\end{aligned}
\end{equation}
Now, given a random sample from the experiment, $(t_{i:N}, ..., t_{n1+n2:N})$, and substituting the true (unknown) c.d.f. for its empirical estimate, the empirical DPD is given by:
\begin{equation}
	\begin{aligned}
	\hat{d}_{\beta}\left(G_T(\cdot), F_T(\cdot|\boldsymbol{\lambda})\right)&=K+\int_{0}^{\tau_1}f_1(t|\lambda_1)^{\beta+1}dt +\int_{\tau_1}^{\tau_2}f_2(t+h|\boldsymbol{\lambda})^{\beta+1}dt+\left(1-F_2(\tau_2+h|\boldsymbol{\lambda})\right)^{\beta+1}
	\\
	&+\frac{\beta+1}{\beta N}\left\{ \sum_{i=1}^{n_1}f_1(t_{i_N}|\lambda_1)^\beta +\sum_{i=n_1+1}^{n_1+n_2}f_2(t_{i_N}+h|\boldsymbol{\lambda})^\beta \right.
	\\
	&\left.+ (N-n_1-n_2)\left(1-F_2(\tau_2+h|\boldsymbol{\lambda})\right)^{\beta}\right\}.
	\end{aligned}
\end{equation}
\par
Remark that all terms depending exclusively on the true distribution can be ignored on the minimization. Then, the MDPDE for a SSALT model and assuming a log-logistic relationship as in \eqref{eq:funcrelation}, with fixed 
tuning parameter $\beta$, \( \beta > 0 \), is defined as:
\begin{equation}
(\hat{a_0}, \hat{a_1})_\beta=arg \min_{a \in \mathbb{R} \times \mathbb{R}^-} H_{N}^{\beta}(a_0, a_1),
\label{eq:dpd_min}
\end{equation}
with:
\begin{equation} \label{eq:Hn}
	\begin{aligned}
		H_{N}^{\beta}(a_0, a_1) =& \int_{0}^{\tau_1} f_1(t|\lambda_1)^{\beta+1}dt 
		+ \int_{\tau_1}^{\tau_2} f_2\left(t||\boldsymbol{\lambda}\right)^{\beta+1}dt 
		+ \left(1-F_2\left(\tau_2 ||\boldsymbol{\lambda}\right)\right)^{\beta+1}
		\\
		&+\frac{\beta+1}{\beta N} \left\{
		  \sum_{i=1}^{n_1} f_1(t_i|\lambda_1)^\beta + 
		  \sum_{i=n_1+1}^{n_1+n_2} f_2\left(t_i||\boldsymbol{\lambda}\right)^\beta \right.
		  \\
		  & \left. + (N-n_1-n_2)\left(1-F_2\left(\tau_2 \, \middle|\, |\boldsymbol{\lambda}\right)\right)^{\beta} 
		 \right\}.
	\end{aligned}
\end{equation}
The family of MDPDE can be extended at $\beta=0$, yielding the MLE defined in \eqref{eq:kl_min}.

Based on \eqref{eq:dpd_min} we have
\begin{equation}
\hat{\lambda}_{i}^{\beta}=\exp\left(\hat{a}_0^{\beta}+\hat{a}_1^{\beta}x_i\right).
\end{equation}
In the next result we present explicit expressions for the DPD-based function $H_{N}^{\beta}(a_0, a_1)$ yielding the MDPDEs.
\begin{proposition} \label{thm:Hn}
The DPD-based loss function of the SSALT model defined in Equation \eqref{eq:Hn}, $H_{N}^{\beta}(a_0, a_1)$. is given under exponential lifetimes by

$$H_{N}^{\beta}(a_0, a_1)=h_1(a_0, a_1)+h_2(a_0, a_1)$$
with
\begin{align*}
	h_1(a_0, a_1)&=\frac{1}{\lambda_1^{\beta}(\beta+1) } - \frac{1}{\lambda_1^{\beta}(\beta+1) } \exp \left( -\frac{\tau_1 }{\lambda_1}(\beta+1) \right) \left(\frac{1}{\lambda_1^\beta}-\frac{1}{\lambda_2^\beta} \right)
	\\
	&+\exp\left(-\frac{\tau_2+h}{\lambda_2}(\beta+1) \right) \left(1 - \frac{1}{\lambda_2^\beta (\beta+1)} \right),
	\\
	h_2(a_0, a_1)&= \frac{\beta+1}{\beta N} \left\{ \frac{1}{\lambda_1^\beta}\sum_{i=1}^{n_1}\exp\left( -\frac{t_i}{\lambda_1}\beta \right) 
	+\frac{1}{\lambda_2^\beta} \sum_{i=n_1+1}^{n_1+n_2} \exp\left(-\frac{t_i+h}{\lambda_2}\beta \right) \right.
	\\
	& + \left. \left(N-n_1-n_2 \right) \exp\left(- \left(\frac{\tau_2+h}{\lambda_2}\beta \right) \right) \right\}	,
\end{align*} \par
being $\lambda_i =\exp\left(a_0+a_1x_i \right)\quad i=1, 2.$

\end{proposition}

\section{The asymptotic distribution of the minimum density power divergence estimator \label{sec:asymptotic}}

In this Section the asymptotic distribution of the MDPDEs obtained from the minimization of the DPD-based function stated in Proposition \ref{thm:Hn} is presented.
\begin{theorem} \label{thm:asympa0}
For the SSALT model under the assumption of exponential lifetimes, with the true regression vector denoted by \(a_0, a_1 \), the asymptotic distribution of the MDPDE of $a_0$, \( \hat{a}_0^\beta \), with fixed tuning parameter $\beta$ is given by
\[
\sqrt{N}\left(\hat{a_0}^\beta-a_0 \right) \xrightarrow[N \to \infty]{\mathscr{L}} \mathcal{N}\left(0, J_{\beta}(a_0)^{-1}K_\beta(a_0)J_\beta(a_0)^{-1} \right).
\]
Being
\[
J_\beta(a_0)=J_{0, \tau_1}^{\beta}(a_0)+J_{\tau_1, \tau_2}^{\beta}(a_0)+J_{\tau_2}^{\beta}(a_0),
\]
with:
\small
\begin{align*}
	&\begin{aligned}
		J^\beta_{0, \tau_1}(a_0)&=\frac{1}{\lambda_1^\beta(\beta+1)}\left(\exp\left(-\frac{\tau_1}{\lambda_1}(\beta+1)\right)\left(-1-\frac{1}{(\beta+1)^2}\left(\left(\frac{\tau_1}{\lambda_1}(\beta+1)\right)^2+\frac{2\tau_1}{\lambda_1}(\beta+1)+2\right) \right.\right.
		\\
		&\left.\left.+\frac{2}{\beta+1}\left(\frac{\tau_1}{\lambda_1}(\beta+1)+1\right)\right)+1+\frac{2}{(\beta+1)^2}-\frac{2}{\beta+1}\right);
	\end{aligned}
	\\
	&\begin{aligned}	
		J_{\tau_2}(a_0)=\left(\frac{\tau_2+h}{\lambda_2}\right)^2\exp\left(-\frac{\tau_2+h}{\lambda_2}(\beta+1)\right),
	\end{aligned}
	\\
	&\begin{aligned}
		J_{\tau_1, \tau_2}^\beta(a_0)&=L^2\frac{1}{\lambda_2^{\beta}(\beta+1)}\exp\left(-\frac{h}{\lambda_2}(\beta+1)\right)\left(\exp\left(-\frac{\tau_1}{\lambda_2}(\beta+1) \right)-\exp\left(-\frac{\tau_2}{\lambda_2}(\beta+1) \right) \right)
		\\
		&+\frac{1}{\lambda_2^{\beta}(\beta+1)^3}\exp\left(-\frac{h}{\lambda_2}(\beta+1)\right)\left( \exp\left(-\frac{\tau_1}{\lambda_2}(\beta+1)\right)\left(\left(\frac{\tau_1}{\lambda_2}(\beta+1)\right)^2+\frac{2\tau_1}{\lambda_2}(\beta+1)+2\right)\right.
		\\
		&\left.-\exp\left(-\frac{\tau_2}{\lambda_2}(\beta+1) \right)\left(\left(\frac{\tau_2}{\lambda_2}(\beta+1)\right)^2+\frac{2\tau_2}{\lambda_2}(\beta+1)+2\right)\right)
		\\
		&+\frac{2L}{\lambda_2^\beta(\beta+1)^2}\exp\left(-\frac{h}{\lambda_2}(\beta+1)\right)\left(\exp\left(-\frac{\tau_1}{\lambda_2}(\beta+1)\right)\left(\frac{\tau_1}{\lambda_2}(\beta+1)+1\right) \right.
		\\
		&-\left.\exp\left(-\frac{\tau_2}{\lambda_2}(\beta+1)\right)\left(\frac{\tau_2}{\lambda_2}(\beta+1)+1\right) \right);
		\\
		L&=-1-\frac{\tau_1}{\lambda_2}+\frac{\tau_1}{\lambda_1}
	\end{aligned}
\end{align*} 
and 
\begin{align*}
K_{\beta}(a_0)&=J_{2\beta}(a_0)-\xi_{\beta}(a_0)^2,
\\
\xi_\beta(a_0)&=\xi_{0, \tau_1}^{\beta}(a_0)+\xi_{\tau_1, \tau_2}^{\beta}(a_0)+\xi_{\tau_2}^{\beta}(a_0)
\end{align*}
with
\begin{align*}
&\begin{aligned}
\xi_{0, \tau_1}^{\beta}(a_0)&=-\frac{1}{\lambda_1^{ \beta}(\beta+1)} \left(1-\exp\left(-\frac{\tau_1}{\lambda_1}(\beta+1) \right) \right)+\frac{1}{\lambda_1^\beta (\beta+1)^2}\left(1-\exp\left(-\frac{\tau_1}{\lambda_1}(\beta+1) \right)\left(\frac{\tau_1}{\lambda_1}(\beta+1)+1\right)\right);
\\
\xi_{\tau_2}^{\beta}(a_0)&=\left(\frac{\tau_2+h}{\lambda_2}\right)\exp\left(-\frac{\tau_2+h}{\lambda_2}(\beta+1)\right)
\\
\xi_{\tau_1, \tau_2}^{\beta}(a_0)&=\frac{L}{\lambda_2^{\beta}(\beta+1)}\exp\left(-\frac{h}{\lambda_2}(\beta+1)\right)\left(\exp\left(-\frac{\tau_1}{\lambda_2}(\beta+1) \right)-\exp\left(-\frac{\tau_2}{\lambda_2}(\beta+1) \right) \right)
\\
&+\frac{1}{\lambda_2^\beta(\beta+1)^2}\exp\left(-\frac{h}{\lambda_2}(\beta+1)\right)\left(\exp\left(-\frac{\tau_1}{\lambda_2}(\beta+1)\right)\left(\frac{\tau_1}{\lambda_2}(\beta+1)+1\right) \right.
\\
&-\left.\exp\left(-\frac{\tau_2}{\lambda_2}(\beta+1)\right)\left(\frac{\tau_2}{\lambda_2}(\beta+1)+1\right) \right).
\end{aligned}
\end{align*} \par
\end{theorem} 
\normalsize
\begin{corollary}
	For the SSALT model under the assumption of exponential lifetimes, with the true regression vector denoted by \(a_0, a_1 \), the asymptotic distribution of the MDPDE of $a_0$, at $\beta = 0$, \( \hat{a}_0^0 \) (corresponding to the MLE) is given by
\[
\sqrt{N}\left(\hat{a}_0^0-a_0 \right) \xrightarrow[N \to \infty]{\mathscr{L}} \mathcal{N}\left(0, J_{0}(a_0)^{-1} \right).
\]
Being
\[
J_0(a_0) = J^0_{0, \tau_1}(a_0) + J^0_{\tau_1, \tau_2}(a_0) + J^0_{\tau_2}(a_0)
\]
with
\small
\begin{align*}
	&\begin{aligned}
		J^0_{0, \tau_1}(a_0) &= 1-\exp\left(-\frac{\tau_1}{\lambda_1}\right) \left(1 + \left(\frac{\tau_1}{\lambda_1}\right)^2\right) ;
	\end{aligned}
	\\
		&\begin{aligned}
		J^0_{\tau_2}(a_0) &= \left(\frac{\tau_2+h}{\lambda_2}\right)^2\exp\left(-\frac{\tau_2+h}{\lambda_2}\right)
	\end{aligned}
	\\
	&\begin{aligned}
		J^0_{\tau_1, \tau_2}(a_0) &= L^2 \exp\left(-\frac{h}{\lambda_2}\right)\left(\exp\left(-\frac{\tau_1}{\lambda_2} \right)-\exp\left(-\frac{\tau_2}{\lambda_2} \right) \right)
		\\
		&+\exp\left(-\frac{h}{\lambda_2}\right)\left( \exp\left(-\frac{\tau_1}{\lambda_2}\right)\left(\left(\frac{\tau_1}{\lambda_2}\right)^2+\frac{2}{\lambda_2}\tau_1+2\right) \right.
		\\
		&-\left.\exp\left(-\frac{\tau_2}{\lambda_2} \right)\left(\left(\frac{\tau_2}{\lambda_2}\right)^2+\frac{2}{\lambda_2}\tau_2+2\right)\right)
		\\
		&+2L \exp\left(-\frac{h}{\lambda_2}\right)\left(\exp\left(-\frac{\tau_1}{\lambda_2}\right)\left(\frac{\tau_1}{\lambda_2}+1\right) -\exp\left(-\frac{\tau_2}{\lambda_2}\right)\left(\frac{\tau_2}{\lambda_2}+1\right) \right);
		\\
		L& = -1-\frac{\tau_1}{\lambda_2}+\frac{\tau_1}{\lambda_1}
	\end{aligned}
\end{align*}
Note that the expression $J_0(a_0)$ is coincides with the Fisher information matrix, $I(a_0)$,
\[
I(a_0) = \mathbb{E} \left[ \left( \frac{\partial}{\partial a_0} \log f_T(T|a_0) \right)^2 \right].
\]
\end{corollary}

\begin{theorem} \label{thm:asympa1}
	For the SSALT model under the assumption of exponential lifetimes, with the true regression vector denoted by \(a_0, a_1 \), the asymptotic distribution of the MDPDE of $a_1$, \( \hat{a}_1^\beta \), with fixed tuning parameter $\beta$ is given by
\[
	\sqrt{N}\left(\hat{a_1}^\beta-a_1 \right) \xrightarrow[N \to \infty]{\mathscr{L}} \mathcal{N}\left(0, J_{\beta}(a_1)^{-1}K_\beta(a_1)J_\beta(a_1)^{-1} \right).
\]
Being
\[
	J_\beta(a_1)=J_{0, \tau_1}^{\beta}(a_1)+J_{\tau_1, \tau_2}^{\beta}(a_1)+J_{\tau_2}^{\beta}(a_1),
\]
with
\small
	\begin{align*}
	&\begin{aligned}
		J^\beta_{0, \tau_1}(a_1)&=\frac{x_1^2}{\lambda_1^\beta(\beta+1)}\left(\exp\left(-\frac{\tau_1(\beta+1)}{\lambda_1}\right)\left(-1-\frac{1}{(\beta+1)^2}\left(\left(\frac{\tau_1(\beta+1)}{\lambda_1}\right)^2+\frac{2(\beta+1)\tau_1}{\lambda_1}+2\right) \right.\right.
		\\
		&\left.\left.+\frac{2}{\beta+1}\left(\frac{\tau_1(\beta+1)}{\lambda_1}+1\right)\right)+1+\frac{2}{(\beta+1)^2}-\frac{2}{\beta+1}\right);
	\end{aligned}
	\\
		&\begin{aligned}
		J_{\tau_2}^\beta(a_1)=\left(\frac{\tau_2}{\lambda_2}x_2+\frac{\tau_1}{\lambda_1}x_1-\frac{\tau_1}{\lambda_2}x_2\right)^2\exp\left(-\frac{\tau_2+h}{\lambda_2}(\beta+1)\right),
	\end{aligned}
	\\
	&\begin{aligned}
		J_{\tau_1, \tau_2}^\beta(a_1)&=\left(L^*\right)^2\frac{1}{\lambda_2^{\beta}(\beta+1)}\exp\left(-\frac{h}{\lambda_2}(\beta+1)\right)\left(\exp\left(-\frac{\tau_1}{\lambda_2}(\beta+1) \right)-\exp\left(-\frac{\tau_2}{\lambda_2}(\beta+1) \right) \right)
		\\
		&+\frac{x_2^2}{\lambda_2^{\beta}(\beta+1)^3}\exp\left(-\frac{h}{\lambda_2}(\beta+1)\right)\left( \exp\left(-\frac{\tau_1}{\lambda_2}(\beta+1)\right)\left(\left(\frac{\tau_1}{\lambda_2}(\beta+1)\right)^2+\frac{2\tau_1}{\lambda_2}(\beta+1)+2\right) \right.
		\\
		&-\left.\exp\left(-\frac{\tau_2}{\lambda_2}(\beta+1) \right)\left(\left(\frac{\tau_2}{\lambda_2}(\beta+1)\right)^2+\frac{2\tau_2}{\lambda_2}(\beta+1)+2\right)\right)
		\\
		&+\frac{2L^*x_2}{\lambda_2^\beta(\beta+1)^2}\exp\left(-\frac{h}{\lambda_2}(\beta+1)\right)\left(\exp\left(-\frac{\tau_1}{\lambda_2}(\beta+1)\right)\left(\frac{\tau_1}{\lambda_2}(\beta+1)+1\right) \right.
		\\
		&-\left.\exp\left(-\frac{\tau_2}{\lambda_2}(\beta+1)\right)\left(\frac{\tau_2}{\lambda_2}(\beta+1)+1\right) \right);
		\\
		L^*&=-x_2-\frac{\tau_1}{\lambda_2}x_2+\frac{\tau_1}{\lambda_1}x_1
	\end{aligned}
		\end{align*}
and
\begin{align*}
	K_{\beta}(a_1)&=J_{2\beta}(a_1)-\xi_{\beta}(a_1)^2 
	\\
	\xi_\beta(a_1)&=\xi_{\tau_1}^{\beta}(a_1)+\xi_{\tau_1, \tau_2}^{\beta}(a_1)+\xi_{\tau_2}^{\beta}(a_1).
\end{align*}
		
 \par
with
\begin{align*}
	&\begin{aligned}
		\xi_{\tau_1}^{\beta}(a_1)&=-\frac{x_1}{\lambda_1^{ \beta}(\beta+1) }\left(1-\exp\left(-\frac{\tau_1}{\lambda_1}(\beta+1) \right) \right)+\frac{x_1}{\lambda_1^{\beta} (\beta+1)^2}\left(1-\exp\left(-\frac{\tau_1}{\lambda_1}(\beta+1) \right)\left(\frac{\tau_1}{\lambda_1}(\beta+1)+1\right)\right);
		\\
		\xi_{\tau_2}^{\beta}(a_1)&=\left(\frac{\tau_2}{\lambda_2}x_2+\frac{\tau_1}{\lambda_1}x_1-\frac{\tau_1}{\lambda_2}x_2\right)\exp\left(-\frac{\tau_2+h}{\lambda_2}(\beta+1)\right).
		\\
	\xi_{\tau_1, \tau_2}^{\beta}(a_1)&=L^*\frac{1}{\lambda_2^{\beta}(\beta+1)}\exp\left(-\frac{h}{\lambda_2}(\beta+1)\right)\left(\exp\left(-\frac{\tau_1}{\lambda_2}(\beta+1) \right)-\exp\left(-\frac{\tau_2}{\lambda_2}(\beta+1) \right) \right);
		\\
		&+\frac{x_2}{\lambda_2^\beta(\beta+1)^2}\exp\left(-\frac{h}{\lambda_2}(\beta+1)\right)\left(\exp\left(-\frac{\tau_1}{\lambda_2}(\beta+1)\right)\left(\frac{\tau_1}{\lambda_2}(\beta+1)+1\right) \right.
		\\
		&-\left.\exp\left(-\frac{\tau_2(\beta+1)}{\lambda_2}\right)\left(\frac{\tau_2}{\lambda_2}(\beta+1)+1\right) \right).
	\end{aligned}
\end{align*} \par
\end{theorem}

\begin{corollary}
		For the SSALT model under the assumption of exponential lifetimes, with the true regression vector denoted by \(a_0, a_1 \), the asymptotic distribution of the MDPDE of $a_1$, at $\beta = 0$, \( \hat{a}_1^0 \) (corresponding to the MLE) is given by
\[
\sqrt{N}\left(\hat{a}_1^0-a_1 \right) \xrightarrow[N \to \infty]{\mathscr{L}} \mathcal{N}\left(0, J_{0}(a_1)^{-1} \right).
\]
Being
\[
J_0(a_1) = J^0_{0, \tau_1}(a_1) + J^0_{\tau_1, \tau_2}(a_1) + J^0_{\tau_2}(a_1)
\]
with
\small
\begin{align*}
	&\begin{aligned}
	  J^0_{0, \tau_1}(a_1) &=x_1^2 \left(1 - \exp\left(-\frac{\tau_1}{\lambda_1}\right) \left(1+\left(\frac{\tau_1}{\lambda_1}\right)^2 \right)\right).
	\end{aligned}
	\\
	&\begin{aligned}
	 J_{0, \tau_2}^0(a_1) &= L^{*2} \exp\left(-\frac{h}{\lambda_2}\right) \left(\exp\left(-\frac{\tau_1}{\lambda_2} \right) - \exp\left(-\frac{\tau_2}{\lambda_2} \right) \right)
		  \\
		  &+x_2^2exp\left(-\frac{h}{\lambda_2}\right) \left( \exp\left(-\frac{\tau_1}{\lambda_2}\right) \left(\left(\frac{\tau_1}{\lambda_2}\right)^2 + \frac{2\tau_1}{\lambda_2} + 2\right) \right. 
		  \\
		  &\left. - \exp\left(-\frac{\tau_2}{\lambda_2} \right) \left(\left(\frac{\tau_2}{\lambda_2}\right)^2 + \frac{2\tau_2}{\lambda_2} + 2\right)\right)
		   \\
		  &+ 2L^*x_2 \exp\left(-\frac{h}{\lambda_2}\right) \left(\exp\left(-\frac{\tau_1}{\lambda_2}\right) \left(\frac{\tau_1}{\lambda_2} + 1\right) - \exp\left(-\frac{\tau_2}{\lambda_2}\right) \left(\frac{\tau_2}{\lambda_2} + 1\right) \right);
		\\
		  L^* &= -x_2 - \frac{\tau_1}{\lambda_2}x_2 + \frac{\tau_1}{\lambda_1}x_1
	\end{aligned}
	\\
	&\begin{aligned}
		  J_{\tau_2}^0(a_1) &= \left(\frac{\tau_2}{\lambda_2}x_2 + \frac{\tau_1}{\lambda_1}x_1 - \frac{\tau_1}{\lambda_2}x_2\right)^2 \exp\left(-\frac{\tau_2 + h}{\lambda_2}\right).
	\end{aligned}
\end{align*}

Again, the expression $J_0(a_1)$ corresponds to the Fisher information, $I(a_1)$, 
\[
I(a_1) = \mathbb{E} \left[ \left( \frac{\partial}{\partial a_1} \log f(T|a_1) \right)^2 \right].
\]
\end{corollary}

\begin{theorem} \label{teo:theo_6}
		For the SSALT model under the assumption of exponential lifetimes and true regression vector denoted by \(a_0, a_1 \), the asymptotic distribution of the MDPDE of the model parameter vector $(a_0, a_1)$, $(\hat{a_0^{\beta}}, \hat{a_1^{\beta}})$, with fixed tuning parameter $\beta$ is given by
\[
	\sqrt{N} \left(\left(\hat{a}_0^\beta, \hat{a}_1^\beta\right)^{\top} - \left(a_0, a_1\right)^{\top}\right)\xrightarrow[N \to \infty]{\mathscr{L}} \mathcal{N}\left(0, \boldsymbol{J}_{\beta}(a_0, a_1)^{-1}\boldsymbol{K}_\beta(a_0, a_1)\boldsymbol{J}_\beta(a_0, a_1)^{-1} \right).
\]
Being
\begin{align*}
	\boldsymbol{J}_\beta(a_0, a_1)&=\begin{pmatrix}		
		J_\beta(a_0) &J_\beta(a_0, a_1)\\
		J_\beta(a_1, a_0) &J_\beta(a_1)
	\end{pmatrix}
	\\
	\boldsymbol{K}_{\beta}(a_0, a_1)&=J_{2\beta}(a_0, a_1)-\xi_{\beta}(a_0, a_1)^\top \xi_{\beta}(a_0, a_1)
\end{align*} \par	

where $J_\beta(a_0)$ y $J_\beta(a_1)$ are defined in Theorems \ref{thm:asympa0} and \ref{thm:asympa1}, and
\begin{align*}
J_\beta(a_0, a_1)
&=J_{0, \tau_1}^{\beta}(a_0, a_1)+J_{\tau_1, \tau_2}^{\beta}(a_0, a_1)+J_{\tau_2}^{\beta}(a_0, a_1).
\end{align*} \par
with
\begin{align*}
J_{0, \tau_1}^{\beta}(a_0, a_1)&=x_1J_{0, \tau_1}^\beta(a_0)
		\\
J_{\tau_1, \tau_2}^{\beta}(a_0, a_1)&=L^* \cdot L\cdot \frac{1}{\lambda_2^{\beta}(\beta+1)}\exp\left(-\frac{h}{\lambda_2}(\beta+1)\right)\left(\exp\left(-\frac{\tau_1}{\lambda_2}(\beta+1) \right)-\exp\left(-\frac{\tau_2}{\lambda_2}(\beta+1) \right) \right) 
\\
&+x_2\frac{1}{\lambda_2^{\beta}(\beta+1)^3}\exp\left(-\frac{h}{\lambda_2}(\beta+1)\right)\left( \exp\left(-\frac{\tau_1}{\lambda_2}(\beta+1)\right)\left(\left(\frac{\tau_1}{\lambda_2}(\beta+1)\right)^2+\frac{2\tau_1}{\lambda_2}(\beta+1)+2\right) \right.
	\\
	&-\left.\exp\left(-\frac{\tau_2}{\lambda_2} (\beta+1)\right)\left(\left(\frac{\tau_2}{\lambda_2}(\beta+1)\right)^2+\frac{2\tau_2}{\lambda_2}(\beta+1)\tau_2+2\right)\right)
	\\
	&+\left(L^*+x_2L\right)\frac{1}{\lambda_2^\beta(\beta+1)^2}\exp\left(-\frac{h}{\lambda_2}(\beta+1)\right)\left(\exp\left(-\frac{\tau_1}{\lambda_2}(\beta+1)\right)\left(\frac{\tau_1}{\lambda_2}(\beta+1)+1\right) \right.
	\\
	&-\left.\exp\left(-\frac{\tau_2}{\lambda_2}(\beta+1)\right)\left(\frac{\tau_2}{\lambda_2}(\beta+1)+1\right) \right)
	\\
	L& = -1-\frac{\tau_1}{\lambda_2}+\frac{\tau_1}{\lambda_1}				
	\\
	L^* &= -x_2 - \frac{\tau_1}{\lambda_2}x_2 + \frac{\tau_1}{\lambda_1}x_1
	\\
	J_{\tau_2}^{\beta}(a_0, a_1)&=\left(\frac{\tau_2}{\lambda_2}x_2+\frac{\tau_1}{\lambda_1}x_1-\frac{\tau_1}{\lambda_2}x_2\right)\left(\frac{\tau_2+h}{\lambda_2}\right)\exp \left(-\left(-\frac{\tau_2+h}{\lambda_2}\right)(1+\beta)\right).
\end{align*} 
and
\[
\boldsymbol{\xi}_{\beta}(a_0, a_1)=\left(\xi_{\beta}(a_0), \xi_{\beta}(a_1)\right).
\]
\end{theorem}

\begin{corollary}
	For the SSALT model under the assumption of exponential lifetimes and true regression vector denoted by \(a_0, a_1 \), the asymptotic distribution of the MDPDE at $\beta=0$ (corresponding with the MLE), $(\hat{a_0^{0}}, \hat{a_1^{0}})$, with fixed tuning parameter $\beta$ is given by
\[
	\sqrt{N} \left(\left(\hat{a}_0^0, \hat{a}_1^0\right)^{\top} - \left(a_0, a_1\right)^{\top}\right)\xrightarrow[N \to \infty]{\mathscr{L}} \mathcal{N}\left(0, \boldsymbol{J}_0(a_0, a_1)^{-1} \right).
\]
Being:
\begin{align*}
	\boldsymbol{J}_0(a_0, a_1)&=\begin{pmatrix}		
		J_0(a_0) &J_0(a_0, a_1)\\
		J_0(a_1, a_0) &J_0(a_1)
	\end{pmatrix}
\end{align*} 	
with:
\small
\begin{align*}
	&\begin{aligned}
	  J^0_{0, \tau_1}(a_0, a_1) &=x_1 \left(1 - \exp\left(-\frac{\tau_1}{\lambda_1}\right) \left(1+\left(\frac{\tau_1}{\lambda_1}\right)^2 \right)\right).
	\end{aligned}
	\\
	&\begin{aligned}
	 J_{0, \tau_2}^0(a_0, a_1) &= L^* \cdot L \exp\left(-\frac{h}{\lambda_2}\right) \left(\exp\left(-\frac{\tau_1}{\lambda_2} \right) - \exp\left(-\frac{\tau_2}{\lambda_2} \right) \right);
		  \\
		  &+x_2exp\left(-\frac{h}{\lambda_2}\right) \left( \exp\left(-\frac{\tau_1}{\lambda_2}\right) \left(\left(\frac{\tau_1}{\lambda_2}\right)^2 + \frac{2\tau_1}{\lambda_2} + 2\right) \right. 
		  \\
		  &\left. - \exp\left(-\frac{\tau_2}{\lambda_2} \right) \left(\left(\frac{\tau_2}{\lambda_2}\right)^2 + \frac{2\tau_2}{\lambda_2} + 2\right)\right)
		   \\
		  &+ \left(L^*+x_2L\right) \exp\left(-\frac{h}{\lambda_2}\right) \left(\exp\left(-\frac{\tau_1}{\lambda_2}\right) \left(\frac{\tau_1}{\lambda_2} + 1\right) - \exp\left(-\frac{\tau_2}{\lambda_2}\right) \left(\frac{\tau_2}{\lambda_2} + 1\right) \right);
		\\
		L& = -1-\frac{\tau_1}{\lambda_2}+\frac{\tau_1}{\lambda_1}			
		\\
		  L^* &= -x_2 - \frac{\tau_1}{\lambda_2}x_2 + \frac{\tau_1}{\lambda_1}x_1
	\end{aligned}
	\\
	&\begin{aligned}
		  J_{\tau_2}^0(a_0, a_1) &= \left(\frac{\tau_2}{\lambda_2}x_2+\frac{\tau_1}{\lambda_1}x_1-\frac{\tau_1}{\lambda_2}x_2\right)\left(\frac{\tau_2+h}{\lambda_2}\right)\exp \left(-\left(-\frac{\tau_2+h}{\lambda_2}\right)\right).
	\end{aligned}
\end{align*}
Thus, the Fisher information, $I(a_0, a_1)$ is given by
\[
\boldsymbol{I}(a_0, a_1) = \mathbb{E} \left[\left( \frac{\partial}{\partial a_0} \log f_T(T|a_0, a_1) \right) \left( \frac{\partial}{\partial a_1} \log f_T(T|a_0, a_1) \right) \right].
\]
\end{corollary}

\section{Lifetime Characteristics Estimation \label{sec:CI}}
Estimates of the model parameters \( a_0 \) and \( a_1 \) have been obtained along with their asymptotic distribution so far.These estimates enable the approximation of the lifetime c.d.f. and associated functions. However, In certain reliability experiments, the primary objective is to estimate some specific characteristics of the lifetime, such as mean lifetime to failure (MTTF), rather than the c.d.f. itself. In the following, we address the robust estimation of key lifetime characteristics based on the proposed MDPDEs and the construction of the corresponding confidence intervals (CIs) using the asymptotic distribution of the estimators and the Delta method.

The reliability of a device upon a mission time $t$ can be determined through the complement of the failure probability. That is, under normal operating conditions the reliability of a device at a mission time $t$ is given by
%
\begin{equation}
R_0(t)=\exp\left(-\frac{t}{\lambda_0}\right)=\exp\left(-t \exp(-a_0-a_1x_0)\right).
\end{equation}
Once the regressions parameters $(a_0, a_1)$ are robustly estimated (and so are the scale parameter, $\lambda_0$), one can estimate the reliability by plugging-in the estimates on the previous expression.
That is, the robust estimate of the reliability at mission time $t$ based on the MDPDE with tuning parameter $\beta, $ is obtained as 
\begin{equation}
	\widehat{R}^\beta_0(t)=\exp\left(-\frac{t}{\widehat{\lambda}^\beta_0}\right)=\exp\left(-t \exp(-\widehat{a}^\beta_0-\widehat{a}^\beta_1x_0)\right).
\end{equation}

The following result states the asymptotic distribution of the robust MDPDE with tuning parameter $\beta$ for the reliability of the device at a fixed mission time $t.$
\begin{theorem}
	Let \( (a_0, a_1)\) be the true parameter of the regression vector and let \( \hat{\textbf{a}}^\beta = (\hat{a}_0^\beta, \hat{a}_1^\beta) \) be the MDPDE with parameter \(\beta\). Then. the asymptotic distribution of the estimated reliability under normal operating conditions based on the MDPDE at a mission time $t$, \( \hat{R}_0^\beta(t) \), is given by
	\[
	\sqrt{N}\left(\hat{R}_0^\beta(t) - R_0(t) \right) \xrightarrow[N \to \infty]{\mathscr{L}} \mathcal{N}\left(0, \sigma_{\beta}\left( R_0(t)\right)^2 \right) .
	\]
	with
	\[
	\sigma_{\beta}\left( R_0(t)\right)^2=\nabla h_R(\boldsymbol{a}_0)^{T}  J_{\beta}(\boldsymbol{a}_0)^{-1}K_\beta(\boldsymbol{a}_0)J_\beta(\boldsymbol{a}_0)^{-1}\nabla h_R(\boldsymbol{a}_0)
	\]
	where $K_\beta(\boldsymbol{a}_0)$ and $J_\beta(\boldsymbol{a}_0)$ are defined in Theorem \ref{teo:theo_6}, and $\nabla h_R(\boldsymbol{a}_0)^{\top}=\left( - R_0(t)\frac{t}{\lambda_0}, - R_0(t) \frac{tx_0}{\lambda_0}\right)$ the gradient of the function $h_R(\boldsymbol{a}) = \exp(-t\exp(a_0+a_1x_0))$ with respect to the model parameters.
	\label{teo:theo_7}
\end{theorem}


Let us now consider the estimation of the time by which more than a proportion $\alpha$ of devices are expected to have failed at normal operating stress level. This quantile estimation relies on the inverse of the reliability function as follows
\begin{equation}
Q_{1-\alpha}=R_0^{-1}(1-\alpha)=-log(1-\alpha)\lambda_0,
\end{equation}
and so, a robust estimator of the $(1-\alpha)$-quantile based on the MDPDE with tuning parameter $\beta$ is given by
\begin{equation} \label{eq:Q1-alpha}
	\widehat{Q}^\beta_{1-\alpha}=\widehat{R}_0^{-1}(1-\alpha)=-\log(1-\alpha)\widehat{\lambda}_0 =-\log(1-\alpha)\exp(\widehat{a}^\beta_0+\widehat{a}^\beta_1x_0), 
\end{equation}

\begin{theorem}
	Given the same conditions as in the Theorem \ref{teo:theo_7}, the asymptotic distribution
	of the estimated $ (1-\alpha )-$ quantile, under normal operating conditions, based on the MDPDE with tuning parameter $\beta$,  \(\hat{Q}_{1-\alpha}^\beta \), is given by
	\[
	\sqrt{N}\left(\hat{Q}_{1-\alpha}^\beta - Q_{1-\alpha} \right) \xrightarrow[N \to \infty]{\mathscr{L}} \mathcal{N}\left(0, \sigma_{\beta}\left( Q_{1-\alpha}\right)^2 \right) .
	\]
	with
	\[
	\sigma_{\beta}\left( Q_{1-\alpha}\right)^2=\nabla h_Q(\boldsymbol{a}_0)^{T}, J_{\beta}(\boldsymbol{a}_0)^{-1}K_\beta(\boldsymbol{a}_0)J_\beta(\boldsymbol{a}_0)^{-1}\nabla h_Q(\boldsymbol{a}_0)
	\]
	where $K_\beta(\boldsymbol{a}_0)$ and $J_\beta(\boldsymbol{a}_0)$ are defined in Theorem \ref{teo:theo_6}, and $\nabla h_Q(\boldsymbol{a})^{\top}=\left( Q_{1-\alpha}, Q_{1-\alpha}x_0\right)$ is the gradient of the function $h_Q(\boldsymbol{a}) = Q_{1-\alpha} = \log\left(1-\alpha\right)\lambda_0.$
	\label{teo:theo_8}
\end{theorem}


Finally, another important and widely used characteristic is the MTTF. For exponential distributions, the expected lifetime of the device is given by
\begin{equation} \label{eq:MTTF}
E_T=\mathbb{E}[T]=\lambda_0=\exp\left(a_0+a_1x_0\right).
\end{equation}


\begin{theorem}
Given the same conditions as in the Theorem \ref{teo:theo_7}, the asymptotic distribution
of the estimated MTTF under normal operating conditions, based on the MDPDE with tuning parameter $\beta$, \(\hat{E}_{T}^\beta \), is given by
\[
\sqrt{N}\left(\hat{E}_T^\beta- E_{T} \right) \xrightarrow[N \to \infty]{\mathscr{L}} \mathcal{N}\left(0, \sigma_{\beta}\left( E_{T} \right)^2 \right) .
\]
with
\[
\sigma_{\beta}\left( E_{T}\right)^2=\nabla h_E(\boldsymbol{a}_0)^{T}, J_{\beta}(\boldsymbol{a}_0)^{-1}K_\beta(\boldsymbol{a}_0)J_\beta(\boldsymbol{a}_0)^{-1}\nabla h_E(\boldsymbol{a}_0)
\]
where $K_\beta(\boldsymbol{a}_0)$ and $J_\beta(\boldsymbol{a}_0)$ are defined in Theorem \ref{teo:theo_6}, and $\nabla h_E(\boldsymbol{a})^{\top}=\left( E_{T}, E_{T}x_0\right)$ is the gradient of the function $h_E(\boldsymbol{a}) = \exp\left(a_0+a_1x_0\right).$
\label{teo:theo_9}
\end{theorem}

Because $\widehat{\boldsymbol{a}}^{\beta}$ is a consistent estimator, we can build an approximated two sided (1-$\alpha$)$\cdot100\%$ CI for the lifetime characteristics robust estimates as follows,
\[
\hat{R}_0^\beta(t) \pm z_{\alpha/2}\frac{\sigma_{\beta}\left( R_0(t)\right)}{\sqrt{N}} \quad \hat{Q}_{1-\alpha}^\beta  \pm z_{\alpha/2}\frac{\sigma_{\beta}\left( Q_{1-\alpha}\right)}{\sqrt{N}} \quad \hat{E}_T^\beta \pm z_{\alpha/2}\frac{\sigma_{\beta}\left( E_{T}\right)}{\sqrt{N}}
\]
where $\sigma_{\beta}\left( R_0(t)\right)$, $\sigma_{\beta}\left( Q_{1-\alpha}\right)$ and $\sigma_{\beta}\left( E_{T}\right)$ where defined in Theorems \ref{teo:theo_7}-\ref{teo:theo_9}.

The approximated confidence intervals derived above were based on the asymptotic properties of the MDPDEs, so they are useful for large sample sizes. In studies with small sample sizes, it may be necessary to truncate the confidence intervals, since some quantities may violate some natural constraints. For example, quantities like the MTTF or distribution quantiles must remain positive, and reliability values must lie within the [0, 1] range. To address this issue, Viveros and Balakrishnan (1993) applied natural transformations to these quantities to ensure that the constraints are satisfied.
This approach allowed for more accurate confidence intervals.
For example, for to construct CI for the reliability, they used a logit transform as
\begin{equation}
\phi=\phi\left(R_0(t)\right)=logit\left(R_0(t)\right)=\log\left(\frac{R_0(t)}{1-R_0(t)}\right).
\end{equation}
The asymptotic distribution and CI for these transformed values can then be derived using once again the Delta method.
As a result, the transformed CI for the reliability, obtained by inverting CI for $\phi$, does not require truncation, making it particularly convenient.
Indeed, this logit transformation is a natural and widely used choice when dealing with parameters that represent probabilities. 

Algebraically inverting the CI for the transformed reliability, the asymptotic $(1-\alpha)\cdot 100\%$ CI for the original reliability measure is obtained as follows
\[
\left[ \frac{\hat{R}_{\beta, 0}(t)}{\hat{R}_{\beta, 0}(t)+\left(1-\hat{R}_{\beta, 0}(t)\right)S}, \frac{\hat{R}_{\beta, 0}(t)}{\hat{R}_{\beta, 0}(t)+\left(1-\hat{R}_{\beta, 0}(t)\right)/S}\right],
\]
with S=$\exp\left(\frac{z_{\alpha/2}}{\sqrt{N}} \frac{\sigma_{\beta}\left( R_0(t)\right)}{\hat{R}_{\beta, 0}(t)+\left(1-\hat{R}_{\beta, 0}(t)\right)}\right)$ and $\sigma_{\beta}\left( R_0(t)\right)$ as defined in Theroem \ref{teo:theo_7}.

A similar approach can be used for the quantiles and MTTF defined in Equations \eqref{eq:Q1-alpha} and \eqref{eq:MTTF}. Since both quantities must be strictly positive, a logarithmic transformation is a natural choice to map the quantities into the real line \( \mathbb{R} \). 
Then, applying the Delta method we can derive the corresponding asymptotic distributions, and subsequently inverting the logarithmic transformations, we obtain the approximated transformed confidence intervals for \( Q_{1 - \alpha} \) and \( E_T\) as follows
\[
\left[ \widehat{Q}^{\beta}_{1-\alpha} \exp\left( -\frac{z_{\alpha/2}}{\sqrt{N}} \frac{\sigma_{\beta}\left( \widehat{Q}_{1-\alpha} \right)}{\widehat{Q}^{\beta}_{1-\alpha}} \right),
    \widehat{Q}^{\beta}_{1-\alpha} \exp\left( \frac{z_{\alpha/2}}{\sqrt{N}} \frac{\sigma_{\beta}\left( \widehat{Q}_{1-\alpha} \right)}{\widehat{Q}^{\beta}_{1-\alpha}} \right)
\right],
\]
and
\[
\left[ \widehat{E_T}^{\beta} \exp\left( -\frac{z_{\alpha/2}}{\sqrt{N}} \frac{\sigma_{\beta}\left( \widehat{E_T} \right)}{\widehat{E_T}^{\beta}} \right),
    \widehat{E_T}^{\beta} \exp\left( \frac{z_{\alpha/2}}{\sqrt{N}} \frac{\sigma_{\beta}\left( \widehat{E_T}\right)}{\widehat{E_T}^{\beta}} \right)
\right],
\]
respectively, with $\sigma_{\beta}\left( \widehat{Q}_{1-\alpha} \right)$ and $\sigma_{\beta}\left( \widehat{E_T} \right)$ were defined in Theorems \ref{teo:theo_8} and \ref{teo:theo_9}.

\section{Simulation study \label{sec:simulation}}

In this section, we analyze the behavior of the MDPDEs, focusing specifically on their robustness in the presence of outliers. To this end, a simulation study is conducted. 

\begin{figure}[h]
	\centering
	\begin{tikzpicture}
		\begin{axis}[
			xlabel={Time},
			ylabel={Stress level},
			width=8cm,
			height=5cm,
			domain=0:33,
			ymin=-0.1, ymax=2.25,
			axis lines=left
			]
			\addplot[
			mark=o,
			thick,
			black,
			const plot
			] coordinates {
				(-1, 0) (0, 0) (0, 1) (10, 1) (10, 2) (33, 2) (33, 0) (33, 0)
			};
		\end{axis}
	\end{tikzpicture}
	\caption{Step-stress test.}
	\label{fig:simulation-step-stress}
\end{figure}
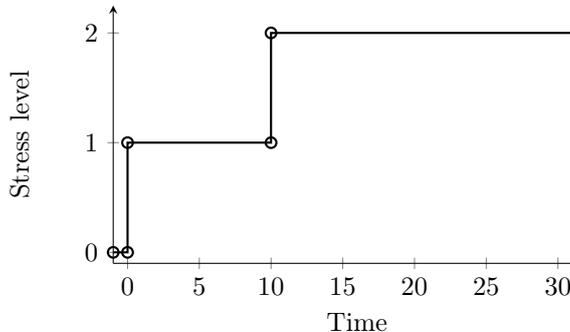

The data were simulated based on an step-stress experimental setup that starts with a stress level of \( x_1 = 1 \). At time \( \tau_1 = 10 \), the stress level increases to \( x_2 = 2 \), and the experiment concludes at \( \tau_2 = 33\). The true parameter values are set to be \( a_0 = 3.5 \) and \( a_1 = -1 \). These parameters values were used in Balakrishnan et al. (2007) for a similar simulated experiment.
Note that \( a_1 \) is negative, so that higher stress levels correspond to shorter expected lifetimes. Thus, we exponential lifetime scale parameters (and MTTF) at constant stress levels $x_1=1$ and $x_2=2$ are 
\[
\lambda_1 = \exp(3.5 - 1) \approx 12.18 \quad \text{and} \quad \lambda_2 = \exp(3.5 - 2) \approx 4.48.
\]
Figure \ref{fig:simulation-step-stress} graphically describes the step-stress design over time.

In this context, an outlier refers to a notable presence of observations located in areas where the underlying p.d.f. is low. Then, to introduce contamination of data, a $\nu$ proportion of the observations is generated from a shifted exponential distribution at 
\(t = 31\), having very short MTTF of 0.5. In other words, an unexpectedly high number of outliers is produced shortly before the experiment ends, corresponding to lifetimes with low probability under the true distribution.

\subsection{Performance for the MSE \label{subsec:perf_MSE}}

We first examine the behavior of the robust estimators for different values of \(\beta\) (including the MLE, \(\beta = 0\)) in the presence of outliers in terms of their estimating error.

To examine the influence of outlying data, we empirically estimate the mean squared error (MSE) of the MDPDEs under several proportions of outliers \( \nu = 0, 0.5\%, \dots, 6\% \) and tuning parameter values \( \beta = 0, 0.2, 0.4, 0.6, 0.8, 1 \).
For a fixed proportion of outliers, a total of $B=4000$ sampled are simulated with a sample size of \( n = 520 \).
For each simulated sample of the step-stress experiment, \( i=1, ..., B \), we compute the MDPDEs of the model parameters, \( \hat{a}_{0, i}^\beta \) and \( \hat{a}_{1, i}^{\beta} \) for \( \beta = 0, 0.2, 0.4, 0.6, 0.8, 1 \). Therefore, for a given outlier proportion \( \nu \), the empirical MSE of the MDPDE with tuning parameter $\beta$ for \( a_0 \) and \( a_1 \) is given by
\[
\text{MSE}_{\beta}^{\nu}(a_0) = \frac{1}{4000} \sum_{i=1}^{4000} \left( \hat{a}_{0, \beta, i}^{\nu} - 3.5 \right)^2 
\quad \text{and} \quad 
\text{MSE}_{\beta}^{\nu}(a_1) = \frac{1}{4000} \sum_{i=1}^{4000} \left( \hat{a}_{1, \beta, i}^{\nu} + 1 \right)^2.
\]
\par
 
\begin{figure}[htb]
  \centering
  \includegraphics[width=1\textwidth]{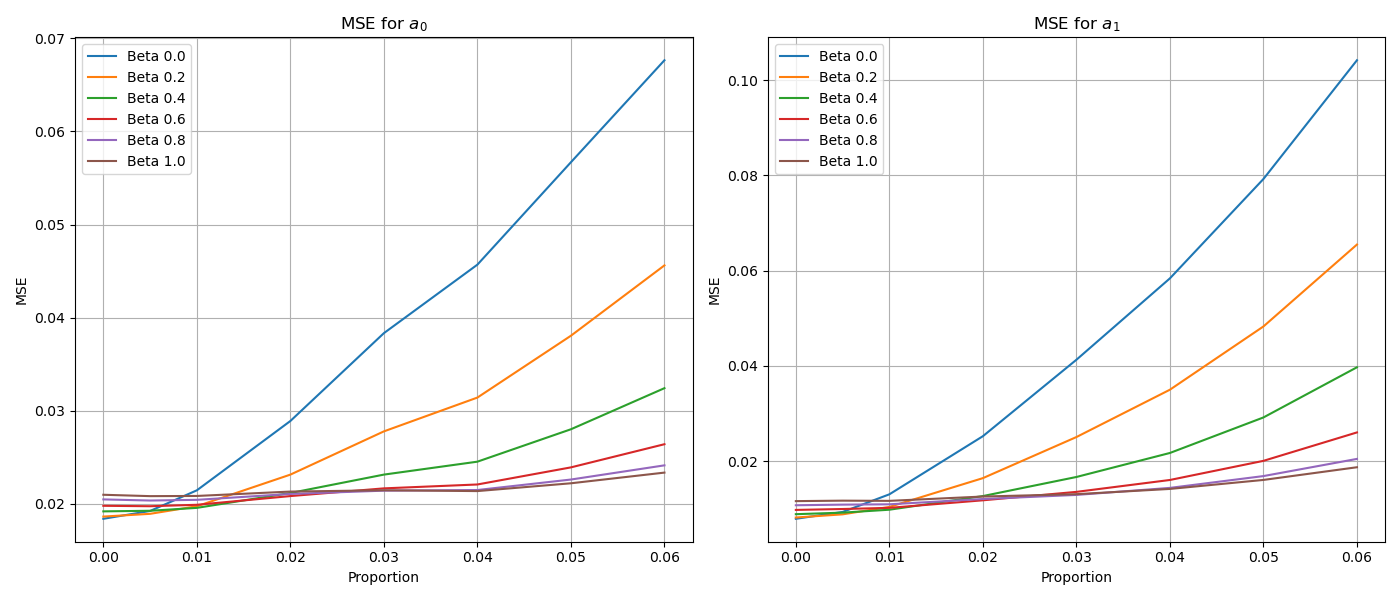}
  \caption{Empirical MSE of the MDPDEs $(\widehat{a}_0^\beta, \widehat{a}_1^\beta)$ for different values of $\beta$.}
  \label{fig:graf-outlier}
\end{figure}

Figure \ref{fig:graf-outlier} shows the empirical MSE of the MDPDEs with different values of the tuning parameter $\beta$ when estimating $a_0$ (left) and $a_1$ (right) under increasing outliers proportion. As shown, the MLE (corresponding to $\beta=0$) outperform the MDPDEs with positive values of $\beta$ in the absence of contamination. However, it rapidly worsens when introducing contamination, and even a $1-2\%$ of outlying observations heavily influence the performance of the MLE. In contrast, the MDPDEs with moderate and high values of the tuning parameter ($\beta$ around 0.4-0.6) perform quite similar than the MLE in the absence of outliers, but keep competitive under contamination in data.

\begin{figure}[h]
  \centering
  \includegraphics[width=1\textwidth]{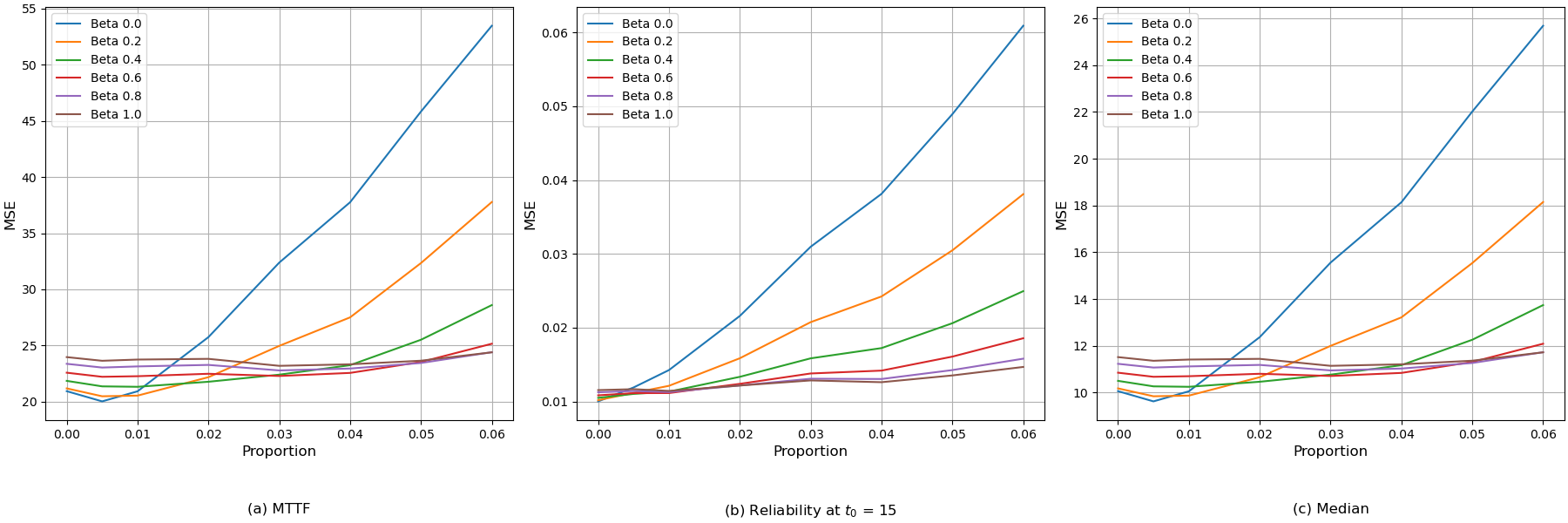}
  \caption{Empirical MSE of the MTTF (left), median (middle) and reliability at $t=14$ (right) estimates based on the MDPDE with different values of $\beta$.}
  \label{fig:graf-outlier_char}
\end{figure}

The robust properties of the MDPDEs are expected to be inherited by the resulting estimates of lifetime characteristics. 
Given the MDPDEs of the model parameters $\hat{a}_0$ and $\hat{a}_1$, estimates of the MTTF ($\hat{\lambda}_0$), the median ($\hat{Q}_{0.5}$), and the reliability $R_0(t)$ at a specific time $t=15$, can directly been computed. 
Figure \ref{fig:graf-outlier_char} presents the estimated MSE of the robust estimates for the MTTF, median and reliability at $t=14$ based on MDPDEs with several values of $\beta.$
As expected, for large values of \(\beta\), the MSE remains controlled across increasing proportions of outliers, while the behavior of the MLE heavily worsens. 

This example illustrates the appealing performance of the MDPDEs with moderate and large values of $\beta$ in the presence of outliers. While estimates based on the MLE are quite heavily influenced by contamination, estimates based on MDPDEs maintain a bounded influence for sufficiently large values of the tuning parameter $\beta$. It is important to emphasize that the objective is not to obtain an estimator completely insensitive to outliers—such insensitivity would typically come at the cost of substantial bias—but rather to limit the influence of outliers in a controlled manner.

\subsection{Performance for the CI}

Finally, let us empirically examine the influence of outlying observations in the estimation of CIs.

A similar simulation set-up as described in Section \ref{sec:simulation} is used.
As before, the true parameter values are set to be \( a_0 = 3.5 \) and \( a_1 = -1 \). 
To reach the nomination level of $\alpha = 0.05$, the sample size is increased to \( n = 10000 \).
Samples are simulated under several number of outliers \( N_{\text{outlier}}= 0, 60, \dots, 160 \) and and the estimation process is repeated over $B=1000$ times. 
For each simulated sample \( i \), the MDPDEs of the model parameters are computed, \( \hat{a}_0^i \) and \( \hat{a}_1^i \) for \( \beta = 0, 0.2, 0.4, 0.6, 0.8, 1 \), along with their corresponding CIs. 

Tables in \ref{tab:results_a0_revisado} present the empirical coverage probability, defined as the proportion of estimated CIs that contain the true parameter value $a_0$, under different outliers proportions. Similar results for $a_1$ are presented in \ref{tab:results_a1_revisado}.
As illustrated, the empirical coverage of the CI based on the different MDPDEs reaches the nominal level in the absence of contamination. However, the coverage of CIs based on the MLE decreases as the level of contamination increases, in contrast to the more stable behavior exhibited by the robust CIs based on MDPDEs with $\beta >0$.It's worth noting the case of $a_1$. With less than 2\% of outliers, the confidence intervals become completely unusable for the MLE, while for certain levels of $\beta$ it maintains some good performance.

\begin{table}[htb]
  \centering
  \caption{Empirical coverage probabilities (Cov), confidence interval widths (CI Width), and mean estimates (Mean) for the estimation of $a_0$ with different values of $\beta$ and Proportion}
  \label{tab:results_a0_revisado}
  \footnotesize
  \begin{subtable}[t]{0.48\textwidth}
    \centering
    \caption{$N_{\text{outlier}}$ = 0}
    \begin{tabular}{lccc}
      \toprule
      $\beta$ & Cov & CI Width & Mean\\
      \midrule
      0.0 & 0.934 & 12.061907 & 3.498804 \\
      0.2 & 0.932 & 12.111462 & 3.498849 \\
      0.4 & 0.937 & 12.235783 & 3.498951 \\
      0.6 & 0.939 & 12.405113 & 3.499061 \\
      0.8 & 0.937 & 12.595768 & 3.499149 \\
      1.0 & 0.938 & 12.789778 & 3.499241 \\
      \bottomrule
    \end{tabular}
    \label{tab:a0_prop0_ajustado_sin_na}
  \end{subtable}%
  \hfill
  \begin{subtable}[t]{0.48\textwidth}
    \centering
    \caption{$N_{\text{outlier}}$ =60}
    \begin{tabular}{lccc}
      \toprule
      $\beta$ & Cov & CI Width & Mean\\
      \midrule
      0.0 & 0.925 & 12.061907 & 3.484228 \\
      0.2 & 0.946 & 12.111462 & 3.488947 \\
      0.4 & 0.950 & 12.235783 & 3.492353 \\
      0.6 & 0.952 & 12.405113 & 3.494690 \\
      0.8 & 0.952 & 12.595768 & 3.496203 \\
      1.0 & 0.951 & 12.789778 & 3.497147 \\
      \bottomrule
    \end{tabular}
    \label{tab:a0_prop0006_ajustado_sin_na}
  \end{subtable}
  
  \bigskip
  
  \begin{subtable}[t]{0.48\textwidth}
    \centering
    \caption{$N_{\text{outlier}}$ = 80}
    \begin{tabular}{lccc}
      \toprule
      $\beta$ & Cov & CI Width & Mean\\
      \midrule
      0.0 & 0.927 & 11.994688 & 3.476483 \\
      0.2 & 0.941 & 12.043532 & 3.482025 \\
      0.4 & 0.948 & 12.164319 & 3.486283 \\
      0.6 & 0.951 & 12.329244 & 3.489505 \\
      0.8 & 0.951 & 12.515250 & 3.491953 \\
      1.0 & 0.951 & 12.714467 & 3.493822 \\
      \bottomrule
    \end{tabular}
    \label{tab:a0_prop0008_ajustado_sin_na}
  \end{subtable}
  \hfill
  \begin{subtable}[t]{0.48\textwidth}
    \centering
    \caption{$N_{\text{outlier}}$ = 100}
    \begin{tabular}{lccc}
      \toprule
      $\beta$ & Cov & CI Width & Mean\\
      \midrule
      0.0 & 0.923 & 11.927469 & 3.468739 \\
      0.2 & 0.933 & 11.975599 & 3.475103 \\
      0.4 & 0.943 & 12.093049 & 3.479986 \\
      0.6 & 0.947 & 12.253375 & 3.483710 \\
      0.8 & 0.948 & 12.434731 & 3.486494 \\
      1.0 & 0.949 & 12.630983 & 3.488616 \\
      \bottomrule
    \end{tabular}
    \label{tab:a0_prop0010_ajustado_sin_na}
  \end{subtable}
  
  \bigskip
  
  \begin{subtable}[t]{0.48\textwidth}
    \centering
    \caption{$N_{\text{outlier}}$ = 120}
    \begin{tabular}{lccc}
      \toprule
      $\beta$ & Cov & CI Width & Mean\\
      \midrule
      0.0 & 0.918 & 11.860250 & 3.460994 \\
      0.2 & 0.928 & 11.907726 & 3.468181 \\
      0.4 & 0.938 & 12.022378 & 3.473709 \\
      0.6 & 0.944 & 12.179506 & 3.478007 \\
      0.8 & 0.946 & 12.357406 & 3.481135 \\
      1.0 & 0.947 & 12.550198 & 3.483509 \\
      \bottomrule
    \end{tabular}
    \label{tab:a0_prop0012_ajustado_sin_na}
  \end{subtable}%
  \hfill
  \begin{subtable}[t]{0.48\textwidth}
    \centering
    \caption{$N_{\text{outlier}}$ = 140}
    \begin{tabular}{lccc}
      \toprule
      $\beta$ & Cov & CI Width & Mean\\
      \midrule
      0.0 & 0.916 & 11.793031 & 3.453250 \\
      0.2 & 0.924 & 11.839853 & 3.461259 \\
      0.4 & 0.933 & 11.951707 & 3.467432 \\
      0.6 & 0.940 & 12.105637 & 3.472304 \\
      0.8 & 0.943 & 12.280081 & 3.475908 \\
      1.0 & 0.944 & 12.470009 & 3.478512 \\
      \bottomrule
    \end{tabular}
    \label{tab:a0_prop0014_ajustado_sin_na}
  \end{subtable}
  
  \bigskip
  
  \begin{subtable}[t]{0.48\textwidth}
    \centering
    \caption{$N_{\text{outlier}}$ = 160}
    \begin{tabular}{lccc}
      \toprule
      $\beta$ & Cov & CI Width & Mean\\
      \midrule
      0.0 & 0.912 & 11.725812 & 3.445506 \\
      0.2 & 0.920 & 11.772004 & 3.454336 \\
      0.4 & 0.929 & 11.881036 & 3.461355 \\
      0.6 & 0.937 & 12.031767 & 3.466801 \\
      0.8 & 0.940 & 12.202755 & 3.470919 \\
      1.0 & 0.942 & 12.389503 & 3.473753 \\
      \bottomrule
    \end{tabular}
    \label{tab:a0_prop0016_ajustado_sin_na}
  \end{subtable}

\end{table}
\begin{table}[htb]
  \centering
  \caption{Empirical coverage probabilities (Cov), confidence interval widths (CI Width), and mean estimates (Mean) for the estimation of $a_1$ with different values of $\beta$ and Proportion}
  \label{tab:results_a1_revisado}
  \footnotesize
  \begin{subtable}[t]{0.48\textwidth}
    \centering
    \caption{ $N_{\text{outlier}}$ = 0}
    \begin{tabular}{lccc}
      \toprule
      $\beta$ & Cov & CI Width & Mean\\
      \midrule
      0.0 & 0.943 & 7.947590 & -0.999390 \\
      0.2 & 0.939 & 8.012745 & -0.999436 \\
      0.4 & 0.935 & 8.195436 & -0.999534 \\
      0.6 & 0.941 & 8.473374 & -0.999651 \\
      0.8 & 0.943 & 8.815907 & -0.999753 \\
      1.0 & 0.936 & 9.186449 & -0.999857 \\
      \bottomrule
    \end{tabular}
    \label{tab:a1_prop0_ajustado_sin_na}
  \end{subtable}%
  \hfill
  \begin{subtable}[t]{0.48\textwidth}
    \centering
    \caption{$N_{\text{outlier}}$ = 60}
    \begin{tabular}{lccc}
      \toprule
      $\beta$ & Cov & CI Width & Mean\\
      \midrule
      0.0 & 0.783 & 7.947590 & -0.975462 \\
      0.2 & 0.857 & 8.012745 & -0.981357 \\
      0.4 & 0.907 & 8.195436 & -0.985887 \\
      0.6 & 0.923 & 8.473374 & -0.989157 \\
      0.8 & 0.942 & 8.815907 & -0.991331 \\
      1.0 & 0.946 & 9.186449 & -0.992655 \\
      \bottomrule
    \end{tabular}
    \label{tab:a1_prop0006_ajustado_sin_na}
  \end{subtable}
  
  \bigskip
  
  \begin{subtable}[t]{0.48\textwidth}
    \centering
    \caption{$N_{\text{outlier}}$ = 80}
    \begin{tabular}{lccc}
      \toprule
      $\beta$ & Cov & CI Width & Mean\\
      \midrule
      0.0 & 0.645 & 7.947590 & -0.967421 \\
      0.2 & 0.784 & 8.012745 & -0.975216 \\
      0.4 & 0.851 & 8.195436 & -0.981213 \\
      0.6 & 0.886 & 8.473374 & -0.985533 \\
      0.8 & 0.913 & 8.815907 & -0.988377 \\
      1.0 & 0.925 & 9.186449 & -0.990062 \\
      \bottomrule
    \end{tabular}
    \label{tab:a1_prop0008_ajustado_sin_na}
  \end{subtable}%
  \hfill
  \begin{subtable}[t]{0.48\textwidth}
    \centering
    \caption{$N_{\text{outlier}}$ = 100}
    \begin{tabular}{lccc}
      \toprule
      $\beta$ & Cov & CI Width & Mean\\
      \midrule
      0.0 & 0.477 & 7.947590 & -0.959675 \\
      0.2 & 0.669 & 8.012745 & -0.969206 \\
      0.4 & 0.794 & 8.195436 & -0.976563 \\
      0.6 & 0.868 & 8.473374 & -0.981875 \\
      0.8 & 0.899 & 8.815907 & -0.985376 \\
      1.0 & 0.914 & 9.186449 & -0.987439 \\
      \bottomrule
    \end{tabular}
    \label{tab:a1_prop0010_ajustado_sin_na}
  \end{subtable}
  
  \bigskip
  
  \begin{subtable}[t]{0.48\textwidth}
    \centering
    \caption{$N_{\text{outlier}}$ = 120}
    \begin{tabular}{lccc}
      \toprule
      $\beta$ & Cov & CI Width & Mean\\
      \midrule
      0.0 & 0.347 & 7.947590 & -0.952546 \\
      0.2 & 0.573 & 8.012745 & -0.963661 \\
      0.4 & 0.736 & 8.195436 & -0.972286 \\
      0.6 & 0.840 & 8.473374 & -0.978548 \\
      0.8 & 0.886 & 8.815907 & -0.982695 \\
      1.0 & 0.904 & 9.186449 & -0.985154 \\
      \bottomrule
    \end{tabular}
    \label{tab:a1_prop0012_ajustado_sin_na}
  \end{subtable}%
  \hfill
  \begin{subtable}[t]{0.48\textwidth}
    \centering
    \caption{$N_{\text{outlier}}$ = 140}
    \begin{tabular}{lccc}
      \toprule
      $\beta$ & Cov & CI Width & Mean\\
      \midrule
      0.0 & 0.221 & 7.947590 & -0.945198 \\
      0.2 & 0.463 & 8.012745 & -0.957965 \\
      0.4 & 0.688 & 8.195436 & -0.967937 \\
      0.6 & 0.802 & 8.473374 & -0.975233 \\
      0.8 & 0.872 & 8.815907 & -0.980105 \\
      1.0 & 0.899 & 9.186449 & -0.983025 \\
      \bottomrule
    \end{tabular}
    \label{tab:a1_prop0014_ajustado_sin_na}
  \end{subtable}
  
  \bigskip
  
  \begin{subtable}[t]{0.48\textwidth}
    \centering
    \caption{$N_{\text{outlier}}$ = 160}
    \begin{tabular}{lccc}
      \toprule
      $\beta$ & Cov & CI Width & Mean\\
      \midrule
      0.0 & 0.139 & 7.947590 & -0.937791 \\
      0.2 & 0.343 & 8.012745 & -0.951994 \\
      0.4 & 0.591 & 8.195436 & -0.963116 \\
      0.6 & 0.731 & 8.473374 & -0.971261 \\
      0.8 & 0.822 & 8.815907 & -0.976694 \\
      1.0 & 0.856 & 9.186449 & -0.979916 \\
      \bottomrule
    \end{tabular}
    \label{tab:a1_prop0016_ajustado_sin_na}
  \end{subtable}

\end{table}

\section{Real-Data Example \label{sec:example}}
This example was studied by Wang and Fei (2003) to obtain the reliability indices of a kind of electronic components at the normal operating temperature of $x_0 = 25^\circ C$. A total of $N = 100$ items from a batch of products were randomly selected for a SSALT with two stress levels $x_1 = 100^\circ C$ and $x_2 = 150^\circ C$. In the original experiment, the stress level was increased when 30 products had failed, and the test continued until 20 more products had failed, resulting in a total of 50 observed failures (Type-II censoring). The observed failure times were as follows:

\begin{itemize}
  \item Failure times at the first stress level $x_1$: 32, 54, 59, 86, 117, 123, 213, 267, 268, 273, 299, 311, 321, 333, 339, 386, 408, 422, 435, 437, 476, 518, 570, 632, 666, 697, 796, 854, 858, 910.
  \item Failure times at the second stress level $x_2$: 16, 19, 21, 36, 37, 63, 70, 75, 83, 95, 100, 106, 110, 113, 116, 135, 136, 149, 172, 186.
\end{itemize}

To illustrate the performance of the MDPDE for the SSALT model under Type-I censoring, we assume that the time of stress change $\tau_1$ is pre-fixed at $t=900$ and that the termination time, $\tau_2$, is pre-fixed at $t=1096$.

The Table~\ref{tab:table_real_data_parameters} shows the parameter estimates under different values of $\beta$, together with their approximate CIs, derived from the asymptotic distributions in Theorems \ref{thm:asympa0} and \ref{thm:asympa1} as follows:
\[
\hat{a}_i \pm \frac{\widehat{\sigma}_\beta(\widehat{a}^\beta_i)}{\sqrt{N}}
\]
with $\widehat{\sigma}_\beta(\widehat{a}^\beta_i) = J_{\beta}(\widehat{a}^\beta_i)^{-1}K_\beta(\widehat{a}^\beta_i)J_\beta(\widehat{a}^\beta_i)^{-1}$ where the corresponding quantities
$J_\beta(a_i)$ and $K_\beta(a_i)$ are defined for $a_0$ in Theorem \ref{thm:asympa0}, and for $a_1$ in Theorem \ref{thm:asympa1}.
Larger values of the tuning parameter leads to wider CIs, but results are quite stable across the different $\beta$s.
\begin{table}[htb]
\centering
\caption{MDPDEs for the electronic components data under different values of $\beta$}
\begin{tabular}{lcccc}
\toprule
$\beta$ & $\hat{a}_0$ & IC($\hat{\theta}_0$) & $\hat{a}_1$ ($\times 10^2$) & IC($\hat{a}_1$) ($\times 10^2$) \\
\midrule
MLE    &10.862 & $[9.476, 12.247]$ & -3.026 & $[-4.159, -1.893]$ \\
0.2    &10.856 & $[9.470, 12.242]$ & -3.021 & $[-4.155, -1.887]$ \\
0.4    & 10.851 & $[9.464, 12.238]$ & -3.017 & $[-4.151, -1.882]$ \\
0.6    & 10.845 & $[9.455, 12.234]$ & -3.012 & $[-4.148, -1.876]$ \\
0.8    &10.837 & $[9.444, 12.230]$ & -3.005 & $[-4.144, -1.867]$ \\
1     &10.832 & $[9.435, 12.229]$ & -3.002 & $[-4.143, -1.860]$ \\
\bottomrule
\end{tabular}
\label{tab:table_real_data_parameters}
\end{table}

As outlined in Section~\ref{sec:CI}, robust estimators of lifetime-related characteristics MTTF, reliability at mission time $t$ and distribution quantiles, can be obtained from the model estimates. 
Robust MTTF estimates based on MDPDEs, along with their corresponding direct and transformed approximate confidence intervals, are presented in Table \ref{tab:table_real_data_values_mean} under three stress levels, $x_0=25$ for normal operating conditions, and $x_1=100$ and $x_2=150$ for the increased stress levels used on the accelerated experiment.
A expected trend is apparent: as the stress level increases ($x_0 < x_1 < x_2$), the estimated mean lifetime decreases significantly, indeed, by $\exp\left(-\hat{a}_1\right).$
Moreover, for constant stress level, the MTTF estimates are relatively stable across different $\beta$ values, thus indicating possibly absence of outlying observations. 
Importantly, under normal operating conditions, the lower bounds of the approximate direct confidence intervals exceed zero, and thus and therefore need to be truncated. In contrast, transformed CIs for the MTTF provide positive interval bounds, which might be more physically meaningful. 
When the stress is increased from $x_1=100^\circ C$ to $x_2=150^\circ C$ the MTTF is reduced to less than $22\%$ times its original value ($78.31\%$ of reduction). 
Generally, a $10^\circ C$ increase in environmental temperature may lead to a reduction in the MTTF of the devices by $(1-\exp(10\widehat{a}_1^\beta ) \cdot 100\%, $ which is approximately of $26 \%.$

\begin{table}[htb]
\centering
\caption{Estimated mean lifetime and asymptotic (direct and transformed) confidence intervals (in hours) of the electronic components under three constant temperatures.}
\begin{tabular}{lcccc}
\toprule
& Mean lifetime & Direct CI & Transformed CI \\
\midrule
\multicolumn{4}{c}{$x_0 = 25$} \\
\midrule
MLE & 6.796 & $[0, 14.335]$ & $[2.241, 20.609]$ \\
0.2 & 6.767 & $[0, 14.278]$ & $[2.231, 20.531]$ \\
0.4 & 6.740 & $[0, 14.228]$ & $[2.219, 20.471]$ \\
0.6 & 6.707 & $[0, 14.171]$ & $[2.204, 20.408]$ \\
0.8 & 6.665 & $[0, 14.099]$ & $[2.185, 20.332]$ \\
1 & 6.640 & $[0, 14.068]$ & $[2.169, 20.323]$ \\
\midrule
\multicolumn{4}{c}{$x_1 = 100$} \\
\midrule
MLE & 0.702 & $[0.452, 0.953]$ & $[0.492, 1.003]$ \\
0.2 & 0.702 & $[0.452, 0.953]$ & $[0.491, 1.003]$ \\
0.4 & 0.702 & $[0.451, 0.952]$ & $[0.491, 1.003]$ \\
0.6 & 0.701 & $[0.450, 0.952]$ & $[0.490, 1.003]$ \\
0.8 & 0.700 & $[0.449, 0.951]$ & $[0.489, 1.002]$ \\
1 & 0.699 & $[0.447, 0.950]$ & $[0.488, 1.002]$ \\
\midrule
\multicolumn{4}{c}{$x_2 = 150$} \\
\midrule
MLE & 0.155 & $[0.087, 0.223]$ & $[0.100, 0.240]$ \\
0.2 & 0.155 & $[0.087, 0.223]$ & $[0.100, 0.241]$ \\
0.4 & 0.155 & $[0.087, 0.224]$ & $[0.100, 0.241]$ \\
0.6 & 0.155 & $[0.087, 0.224]$ & $[0.100, 0.242]$ \\
0.8 & 0.156 & $[0.087, 0.225]$ & $[0.100, 0.242]$ \\
1 & 0.156 & $[0.087, 0.225]$ & $[0.100, 0.243]$ \\
\bottomrule
\end{tabular}
\label{tab:table_real_data_values_mean}
\end{table}

Another quantity of interest in industry is the reliability of the device at a given mission time under normal operating conditions, as well as how this reliability is affected by increases in environmental temperature. Let us fix a mission time $t=600s.$ The MDPDEs for the reliability and their corresponding direct and transformed approximate CIs with different values of $\beta$ are presented in Table \ref{tab:table_real_data_values_survive} for constant temperatures. 
Notably, reliability decreases substantially as the stress level increases, indicating that the product degrades more rapidly at higher temperatures.
Furthermore, the choice of the tuning parameter $\beta$ has a relatively minor impact on the estimated reliability, as all estimates remain quite similar. As for the MTTF, the bounds of the direct CIs under normal operating conditions are truncated to satisfy the natural constraint of lying between 0 and 1. Moreover, the direct and transformed approximate confidence intervals provide similar bounds at low temperature levels, but differ slightly at higher temperatures.
\begin{table}[htb]
\centering
\caption{Estimated reliability at $t=600$ and asymptotic (direct and transformed) confidence intervals (in hours) for the electronic components under three constant temperatures.}
\begin{tabular}{lcccc}
\toprule
& $\hat{R}$(600) & Direct CI & Transformed CI \\
\midrule
\multicolumn{4}{c}{$x_0 = 25$} \\
\midrule
MLE & 0.976 & $[0.949, 1.000]$ & $[0.950, 1.000]$ \\
0.2 & 0.976 & $[0.949, 1.000]$ & $[0.949, 1.000]$ \\
0.4 & 0.976 & $[0.949, 1.000]$ & $[0.949, 1.000]$ \\
0.6 & 0.975 & $[0.948, 1.000]$ & $[0.949, 1.000]$ \\
0.8 & 0.975 & $[0.948, 1.000]$ & $[0.948, 1.000]$ \\
1 & 0.975 & $[0.948, 1.000]$ & $[0.948, 1.000]$ \\
\midrule
\multicolumn{4}{c}{$x_1 = 100$} \\
\midrule
MLE & 0.789 & $[0.722, 0.855]$ & $[0.725, 0.858]$ \\
0.2 & 0.789 & $[0.722, 0.855]$ & $[0.725, 0.858]$ \\
0.4 & 0.789 & $[0.722, 0.856]$ & $[0.724, 0.858]$ \\
0.6 & 0.788 & $[0.721, 0.855]$ & $[0.724, 0.858]$ \\
0.8 & 0.788 & $[0.721, 0.855]$ & $[0.723, 0.858]$ \\
1 & 0.788 & $[0.720, 0.855]$ & $[0.723, 0.858]$ \\
\midrule
\multicolumn{4}{c}{$x_2 = 150$} \\
\midrule
MLE & 0.340 & $[0.179, 0.502]$ & $[0.212, 0.547]$ \\
0.2 & 0.341 & $[0.179, 0.503]$ & $[0.212, 0.548]$ \\
0.4 & 0.342 & $[0.180, 0.504]$ & $[0.213, 0.549]$ \\
0.6 & 0.342 & $[0.180, 0.504]$ & $[0.213, 0.550]$ \\
0.8 & 0.343 & $[0.181, 0.505]$ & $[0.214, 0.550]$ \\
1 & 0.343 & $[0.181, 0.506]$ & $[0.214, 0.551]$ \\
\bottomrule
\end{tabular}
\label{tab:table_real_data_values_survive}
\end{table}

Finally, let us discuss results for the $0.9-$quantile estimates. The $90\%-$quantile represents the time by which $90\%$ of the units are expected to have survived under constant temperature conditions.
Table \ref{tab:table_real_data_values_decile} presents the MDPDEs for the $0.9-$quantile and their corresponding approximate CIs. 
As expected, all $0.9-$quantile estimates show a strong inverse relationship with the temperature; The time at which $90\%$ of the devices are still functioning is much shorter at higher temperatures.
When comparing direct and transformed approximated CIs for the 0.9-quantile, the transformed CIs tend to have higher upper and lower bounds. At normal operating conditions $x_0=25, $ lower bounds of direct CIs resulted negative, and so were truncated. 

\begin{table}[htb]
	\centering
	\caption{MDPDEs for the $90\%-$quantile (in minutes) and their corresponding asymptotic (direct and transformed) confidence intervals for the electronic components under three constant temperatures.}
	\begin{tabular}{lcccc}
		\toprule
		& $\hat{Q}_{0.9}$ & Direct CI & Transformed CI \\
	\midrule
	\multicolumn{4}{c}{$x_0 = 25$} \\
	\midrule
	MLE & 42.960 & $[0, 90.600]$ & $[14.160, 130.260]$ \\
	0.2 & 42.780 & $[0, 90.240]$ & $[14.100, 129.780]$ \\
	0.4 & 42.600 & $[0, 89.940]$ & $[14.040, 129.420]$ \\
	0.6 & 42.420 & $[0, 89.580]$ & $[13.920, 129.000]$ \\
	0.8 & 42.120 & $[0, 89.100]$ & $[13.800, 128.520]$ \\
	1 & 42.000 & $[0, 88.920]$ & $[13.740, 128.460]$ \\
	\midrule
	\multicolumn{4}{c}{$x_1 = 100$} \\
	\midrule
	MLE & 4.440 & $[2.856, 6.024]$ & $[3.108, 6.342]$ \\
	0.2 & 4.440 & $[2.856, 6.024]$ & $[3.108, 6.342]$ \\
	0.4 & 4.434 & $[2.850, 6.018]$ & $[3.102, 6.342]$ \\
	0.6 & 4.428 & $[2.844, 6.018]$ & $[3.096, 6.336]$ \\
	0.8 & 4.422 & $[2.838, 6.012]$ & $[3.090, 6.330]$ \\
	1 & 4.416 & $[2.826, 6.006]$ & $[3.084, 6.330]$ \\
	\midrule
	\multicolumn{4}{c}{$x_2 = 150$} \\
	\midrule
	MLE & 0.978 & $[0.546, 1.410]$ & $[0.630, 1.518]$ \\
	0.2 & 0.978 & $[0.546, 1.410]$ & $[0.630, 1.524]$ \\
	0.4 & 0.984 & $[0.546, 1.416]$ & $[0.630, 1.524]$ \\
	0.6 & 0.984 & $[0.546, 1.416]$ & $[0.630, 1.530]$ \\
	0.8 & 0.984 & $[0.546, 1.422]$ & $[0.630, 1.530]$ \\
	1 & 0.984 & $[0.552, 1.422]$ & $[0.630, 1.536]$ \\
	\bottomrule
	\end{tabular}
	\label{tab:table_real_data_values_decile}
\end{table}

In Balakrishnan et al. (2023b), the authors performed estimations using the same dataset but instead of continuous monitoring, they used interval-censored data, by counting the number of failures at some inspection times in between. The estimated values of the model parameters under both continuous and censored setups are quite similar, but two main differences can be highlighted: Firstly, in the continuous monitoring case, the estimated parameter values become more stable as $\beta$ increases, showing a slower decreasing trend and thus indicating a reduced sensitivity to the choice of $\beta$ in the absence of contamination. Secondly, the CIs obtained under continuous monitoring for a $95\%$ confidence level have are narrower compared to those obtained under censored data.


\section{Conclusions and Future Works \label{sec:concluding}}

In this paper, we have proposed a novel method based on the DPD for robust parameter estimation for SSALTs under continuous monitoring.
For such a purpose, we have defined the DPD for a mixed distribution, used to model the observable c.d.f. of the device's lifetime during the experiment.
The formulation provided in this paper raises various questions that are under simultaneous investigation, such as the derivation of the Influence Function (IF) of the resulting MDPDEs, which is widely used to theoretically determine the robustness properties of an estimator.

Moreover, we have assumed the simplest distribution modelling lifetime data. However, other (more complex) parametric families such as Weibull, log-normal, gamma, log-logistic, Lindley, etc. need to be studied under the mixed distribution formulation. In particular, explicit expressions for the DPD-based loss function and MDPDEs asymptotic variance-covariance need to be obtained.

Moreover, robust inference extends beyond point and interval estimation to include the construction of test statistics designed to mitigate the influence of outliers. For such a purpose, robust Wald-type and Rao-type test statistics based on the robust MDPDEs under the previous mentioned parametric families need to be defined. For the generalized family of Rao-type test statistics, the restricted MDPDEs under mixed distributions must be jointly derived.

On the other hand, throughout this paper we have assumed the CE model to relating the life distribution to the applied stress level, which is equivalent to other life-stress models under exponential lifetimes (see discussion in Section \ref{sec:intro}). However, there exist two other important proposals in the literature to this end, namely the tampered random variable model (TRV) and the tampered failure rate model (TFR), yielding to different expression of the c.d.f. under the assumption of other parametric families, such as the previously mentioned. Thus, MDPDEs under these life-stress models need to be developed and studied.
One last interesting open problem we would like to mention is the formulation of all the previous statistical models, but considering type- II censoring in data.

These research lines are currently being explored, with the goal of defining a new research agenda upon resolving the identified challenges.

\section*{Acknowledgements}
This work was supported by the Spanish Grant PID2021-124933NB-I00.
M. Jaenada and L. Pardo are members of the Interdisciplinary Mathematics Institute (IMI).

\newpage
\appendix
\section*{Appendix}

\section{Proof of Proposition 1}
\begin{proof}
	The expression \( h_1(a_0, a_1) \) is formed by the sum of three elements, let's calculate each one:
	\begin{align*}
		& \begin{aligned}
			\bullet \int_0^{\tau_1} \frac{1}{\lambda_1^{\beta+1}} \exp\left( -\frac{t}{\lambda_1}(\beta+1)\right)dt & = \frac{1}{\lambda_1^{\beta+1}} \int_0^{\tau_1} \exp\left(-\frac{t}{\lambda_1}(\beta+1)\right) dt=\frac{1}{\lambda_1^{\beta+1}} \Bigg[ \frac{\lambda_1}{\beta+1} \exp\left(-\frac{t}{\lambda_1}(\beta+1)\right) \Bigg]_0^{\tau_1} 
			\\
			&= \frac{1}{\lambda_1^{\beta}(\beta+1)}\left(1-\exp \left(-\frac{\tau_1}{\lambda_1}(\beta+1) \right) \right).
		\end{aligned} \\
		& \begin{aligned}
			\bullet \int_{\tau_1}^{\tau_2}\frac{1}{\lambda_2^{\beta+1}}\exp\left(-\frac{t+h}{\lambda_2}(\beta+1)\right)dt &= \frac{1}{\lambda_2^{\beta+1}}\exp\left(-\frac{h}{\lambda_2}(\beta+1) \right) \int_{\tau_1}^{\tau_2}\exp\left(-\frac{t}{\lambda_2}(\beta+1)\right)dt
			\\
			&= \frac{1}{\lambda_2^{\beta+1}}\exp\left((\beta+1)\left(-\frac{\tau_1}{\lambda_1}+\frac{\tau_1}{\lambda_2} \right) \right) \Bigg[\frac{\lambda_2}{\beta+1}\exp\left(-\frac{t}{\lambda_2}(\beta+1)\right) \Bigg]_{\tau_1}^{\tau_2}
			\\
			& = \frac{1}{\lambda_2^{\beta}(\beta+1)}\exp\left(-\left(\frac{1}{\lambda_1}-\frac{1}{\lambda_2} \right)\tau_1(\beta+1) \right)
			\\
			& \quad \cdot \left(\exp\left(-\frac{\tau_1}{\lambda_2}(\beta+1)\right)-\exp\left(-\frac{\tau_2}{\lambda_2}(\beta+1)\right) \right)
			\\
			&= \frac{1}{\lambda_2^\beta(\beta+1)}\left(\exp\left(-\frac{\tau_1}{\lambda_1}(\beta+1)\right)-\exp\left(-\frac{\tau_2+h}{\lambda_2}(\beta+1)\right)\right).
		\end{aligned} \\
		& \begin{aligned}
			\bullet & \left(1-F_2\left(\tau_2+h\right)\right)^{\beta+1} =\exp\left(- \frac{\tau_2+h}{\lambda_2} (\beta+1)\right).
		\end{aligned}
	\end{align*} \par

	Therefore, the expression for \( h_1(a_0, a_1) \) is:
	\begin{align*}
		h_1(a_0, a_1)&=\frac{1}{\lambda_1^{\beta}(\beta+1) } - \frac{1}{\beta+1 }\left(\frac{1}{\lambda_1^\beta}-\frac{1}{\lambda_2^\beta} \right) exp \left( -\frac{\tau_1 }{\lambda_1}(\beta+1)\right) 
		\\
		&+exp\left(-\frac{\tau_2+h}{\lambda_2} (\beta+1) \right) \left(1 - \frac{1}{\lambda_2^\beta (\beta+1)} \right).
	\end{align*} \par
	On the other hand, the expression for \( h_2(a_0, a_1) \) is exactly:
	\begin{align*}
		h_2(a_0, a_1)=& \frac{\beta+1}{\beta N} \left\{ \frac{1}{\lambda_1^\beta}\sum_{i=1}^{n_1}exp\left( -\frac{t_i}{\lambda_1}\beta \right) 
		+\frac{1}{\lambda_2^\beta} \sum_{i=n_1+1}^{n_1+n_2} exp\left(-\frac{t_i+h}{\lambda_2}\beta \right) \right.
		\\
		& + \left. \left(N-n_1-n_2 \right) exp\left(- \frac{\tau_2+h}{\lambda_2}\beta \right) \right\}	.
	\end{align*} \par
\end{proof}

\small

\section{Proof of Theorem 2}

\begin{proof}
	\[
	J_\beta(a_0)=J_{0, \tau_1}^{\beta}(a_0)+J_{\tau_1, \tau_2}^{\beta}(a_0)+J_{\tau_2}^{\beta}(a_0),
	\]
	\begin{align*}
		J_\beta(a_0)&=\int_0^{\tau_1} \left(\frac{\partial \log\left(f_1(t|\lambda_1)\right)}{\partial a_0}\right)^2f_1(t|\lambda_1)^{\beta+1}dt
		\\
		&+\int_{\tau_1}^{\tau_2} \left(\frac{\partial \log\left(f_2(t+h|\boldsymbol{\lambda})\right)}{\partial a_0}\right)^2f_2(t+h|\boldsymbol{\lambda})^{\beta+1}dt
		\\
		&+ \left(\frac{\partial}{\partial a_0}\log\left(1-F_2\left(\tau_2|\boldsymbol{\lambda}\right)\right)\right)^2\left(1-F_2\left(\tau_2|\boldsymbol{\lambda}\right) \right)^{\beta+1}
		\\
		&=J_{0, \tau_1}^{\beta}(a_0)+J_{\tau_1, \tau_2}^{\beta}(a_0)+J_{\tau_2}^{\beta}(a_0).
	\end{align*} \par
	\begin{align*}
		J_{0, \tau_1}^{\beta}(a_0)&=\int_0^{\tau_1} \left(\frac{\partial \log\left(f_1(t|\lambda_1)\right)}{\partial a_0}\right)^2f_1(t|\lambda_1)^{\beta+1}dt,
		\\
		J_{\tau_1, \tau_2}^{\beta}(a_0)&=\int_{\tau_1}^{\tau_2} \left(\frac{\partial \log\left(f_2(t+h|\boldsymbol{\lambda})\right)}{\partial a_0}\right)^2f_2(t+h|\boldsymbol{\lambda})^{\beta+1}dt,
		\\
		J_{\tau_2}^{\beta}(a_0)&= \left(\frac{\partial}{\partial a_0}\log\left(1-F_2\left(\tau_2|\boldsymbol{\lambda}\right)\right)\right)^2\left(1-F_2\left(\tau_2|\boldsymbol{\lambda}\right) \right)^{\beta+1}.
	\end{align*} \par
	\par
	Now the explicit expressions for each term will be obtained:
	\par
	For $J_{0, \tau_1}^{\beta}(a_0)$, we have
	\begin{align*}
		f_1(t|\lambda_1)&=\frac{1}{\lambda_1}\exp\left(-\frac{t}{\lambda_1} \right) \Rightarrow \quad \log\left(f_1(t|\lambda_1)\right) = -\log(\lambda_1) -\frac{t}{\lambda_1} ; \lambda_i=\exp\left(a_0+a_1x_i \right).
		\\
		\text{Then } \quad &\frac{\partial \log\left(f_1(t|\lambda_1)\right)}{\partial a_0}=\frac{\partial \log\left(f_1(t|\lambda_1)\right)}{\partial \lambda_1}\cdot \frac{\partial \lambda_1}{\partial a_0} = \left(-\frac{1}{\lambda_1}+\frac{t}{\lambda_1^2} \right) \lambda_1
		\\
		&=-1+\frac{t}{\lambda_1} \Rightarrow \left(\frac{\partial  \log\left(f_1(t|\lambda_1)\right)}{\partial a_0} \right)^2=1+\frac{t^2}{\lambda_1^2}-2\frac{t}{\lambda_1}.
	\end{align*} \par
	Therefore, we have:
	\begin{align*}
		J_{0, \tau_1}^\beta(a_0)&=\int _0 ^{\tau_1} \left(\frac{\partial  \log\left(f_1(t|\lambda_1)\right)}{\partial a_0} \right)^2 f_1(t|\lambda_1)^{\beta+1} dt
		\\
		&=\int _0 ^{\tau_1} f_1(t|\lambda_1)^{\beta+1} dt+\frac{1}{\lambda_1^2} \int _0 ^{\tau_1} t^2 f_1(t|\lambda_1)^{\beta+1} dt-\frac{2}{\lambda_1}\int _0 ^{\tau_1}t f_1(t|\lambda_1)^{\beta+1} dt
		\\
		&=\int _0 ^{\tau_1} \frac{1}{\lambda_1^{\beta+1}} \exp\left(-\frac{t}{\lambda_1}(\beta+1)\right)dt+\frac{1}{\lambda_1^2} \int _0 ^{\tau_1} \frac{1}{\lambda_1^{\beta+1}} t^2\exp\left(-\frac{t}{\lambda_1}(\beta+1)\right)dt
		\\
		&-\frac{2}{\lambda_1}\int _0 ^{\tau_1} \frac{1}{\lambda_1^{\beta+1}} t \exp\left(-\frac{t}{\lambda_1}(\beta+1)\right) dt=L_1+L_2+L_3.
	\end{align*} \par
	\textit{Note}
	\begin{align*}
		\int t^2\exp\left( -tc \right)dt&=-\frac{1}{c^3} \exp\left(-ct \right) \left(c^2t^2+2ct+2\right)+C,
		\\
		\int t \exp \left(-tc \right)dt& = -\frac{1}{c^2} \exp \left(-ct \right) \left( ct +1 \right)+C.
	\end{align*} \par
	And therefore, we have:
	\begin{align*}
		L_1&=\frac{1}{\lambda_1^{\beta+1}}\int _0 ^{\tau_1} \exp\left(-\frac{t}{\lambda_1}(\beta+1)\right)dt=-\frac{1}{\lambda_1^{\beta+1}}\frac{\lambda_1}{\beta+1} \left[ \exp\left(-\frac{t}{\lambda_1}(\beta+1)\right)\right]_0^{\tau_1}=\frac{1}{\lambda_1^{ \beta}(\beta+1)} \left(1-\exp\left(-\frac{\tau_1}{\lambda_1}(\beta+1) \right) \right);
		\\
		L_2&=\frac{1}{\lambda_1^{\beta+3}} \int _0 ^{\tau_1} t^2\exp\left(-\frac{t}{\lambda_1}(\beta+1)\right)dt=-\frac{1}{\lambda_1^{\beta+3}}\frac{\lambda_1^3}{(\beta+1)^3}\left[\exp\left(-\frac{t}{\lambda_1}(\beta+1) \right)\left(\left(\frac{t}{\lambda_1}(\beta+1) \right)^2+2\frac{t}{\lambda_1}(\beta+1)+2 \right)\right]_0^{\tau_1}
		\\
		&=\frac{1}{\lambda_1^{\beta}(\beta+1)^3}\left(2-\exp\left(-\frac{\tau_1}{\lambda_1}(\beta+1) \right)\left(\left(\frac{\tau_1}{\lambda_1(\beta+1)}\right)^2+\frac{2\tau_1}{\lambda_1}(\beta+1)+2\right)\right);
		\\
		L_3&=-\frac{2}{\lambda_1}\int _0 ^{\tau_1} \frac{1}{\lambda_1^{\beta+1}} t \exp\left(-\frac{t}{\lambda_1}(\beta+1)\right) dt=\frac{2}{\lambda_1^{\beta+1}}\frac{\lambda_1^2}{(\beta+1)^2}\left[\exp\left(-\frac{t}{\lambda_1}(\beta+1)\right)\left(\frac{t}{\lambda_1}(\beta+1)+1 \right) \right]_0^{\tau_1}
		\\
		&=-\frac{2}{\lambda_1^\beta (\beta+1)^2}\left(1-\exp\left(-\frac{\tau_1}{\lambda_1}(\beta+1) \right)\left(\frac{\tau_1}{\lambda_1}(\beta+1)+1\right)\right).
	\end{align*} \par
	Adding the calculated terms, we obtain:
	\begin{align*}
		J_{0, \tau_1}^{\beta}(a_0)=&\exp\left(-\frac{\tau_1}{\lambda_1}(\beta+1)\right) \left( -\frac{1}{\lambda_1^\beta(\beta+1)}-\frac{1}{\lambda_1^\beta(\beta+1)^3}\left(\left(\frac{\tau_1}{\lambda_1}(\beta+1)\right)^2+\frac{2\tau_1}{\lambda_1}(\beta+1)+2\right) \right.
		\\
		&+\left. \frac{2}{\lambda^\beta(\beta+1)^2}\left(\frac{\tau_1}{\lambda_1}(\beta+1)+1 \right) \right)+\frac{1}{\lambda_1^\beta(\beta+1)}+\frac{2}{\lambda_1^\beta(\beta+1)^3}-\frac{2}{\lambda_1^\beta(\beta+1)^2}
		\\
		&=\frac{1}{\lambda_1^\beta(\beta+1)}\left(\exp\left(-\frac{\tau_1}{\lambda_1}(\beta+1)\right)\left(-1-\frac{1}{(\beta+1)^2}\left(\left(\frac{\tau_1}{\lambda_1}(\beta+1)\right)^2+\frac{2\tau_1}{\lambda_1}(\beta+1)+2\right) \right.\right.
		\\
		&\left.\left.+\frac{2}{\beta+1}\left(\frac{\tau_1}{\lambda_1}(\beta+1)+1\right)\right)+1+\frac{2}{(\beta+1)^2}-\frac{2}{\beta+1}\right).
	\end{align*} \par
	For \( J_{\tau_1, \tau_2}(a_0) \), we have:
	\begin{align*}
		f_2(t+h|\boldsymbol{\lambda})&=\frac{1}{\lambda_2}\exp\left(-\frac{1}{\lambda_2}\left(t+\frac{\lambda_1}{\lambda_2}\tau_1-\tau_1\right) \right) \Rightarrow \quad \log\left(f_2(t+h|\boldsymbol{\lambda})\right) = -\log(\lambda_2) -\frac{t}{\lambda_2}-\frac{\tau_1}{\lambda_1}+\frac{\tau_1}{\lambda_2} ;
		\\
		\text{with } \lambda_i&=\exp\left(a_0+a_1x_i \right);
		\\
		\text{Then} \quad \frac{\partial \log\left(f_2(t+h|\boldsymbol{\lambda})\right)}{\partial a_0}&=\frac{\partial \log\left(f_2(t+h|\boldsymbol{\lambda})\right)}{\partial \lambda_1}\cdot \frac{\partial \lambda_1}{\partial a_0}+\frac{\partial \log\left(f_2(t+h|\boldsymbol{\lambda})\right)}{\partial \lambda_2}\cdot \frac{\partial \lambda_2}{\partial a_0} 
		\\
		& = -\frac{1}{\lambda_2}\lambda_2+\frac{t}{\lambda_2^2}\lambda_2-\frac{\tau_1}{\lambda_1^2}\lambda_1+\frac{\tau_1}{\lambda_2^2}\lambda_2=\left(-1-\frac{\tau_1}{\lambda_2}+\frac{\tau_1}{\lambda_1} \right) +\frac{t}{\lambda_2}=L+\frac{t}{\lambda_2};
		\\
		& \Rightarrow \left(\frac{\partial  \log\left(f_2(t+h|\boldsymbol{\lambda})\right)}{\partial a_0} \right)^2=L^2+\frac{1}{\lambda_2^2}t^2+\frac{2L}{\lambda_2}t.
	\end{align*} \par
	Therefore, we have:
	\begin{align*}
		J_{\tau_1, \tau_2}^\beta(a_0)&=\int _{\tau_1} ^{\tau_2} \left(\frac{\partial  \log\left(f_2(t+h|\boldsymbol{\lambda})\right)}{\partial a_0} \right)^2 f_2(t+h|\boldsymbol{\lambda})^{\beta+1} dt
		\\
		&=L^2\int _{\tau_1}^{\tau_2} \frac{1}{\lambda_2^{\beta+1}} \exp \left(-\frac{t+h}{\lambda_2} \left( \beta+1\right)\right)dt+\frac{1}{\lambda_2^2}\int _{\tau_1}^{\tau_2} t^2\frac{1}{\lambda_2^{\beta+1}} \exp \left(-\frac{t+h}{\lambda_2} \left( \beta+1\right)\right)dt
		\\
		&+\frac{2L}{\lambda_2}\int _{\tau_1}^{\tau_2} \frac{t}{\lambda_2^{\beta+1}} \exp \left(-\frac{t+h}{\lambda_2} \left(\beta+1\right)\right)dt
		\\
		&=M_1+M_2+M_3.
		\\
		M_1&=L^2\frac{1}{\lambda_2^{\beta+1}}\exp\left(-\frac{h}{\lambda_2}(\beta+1)\right)\int_{\tau_1}^{\tau_2}\exp\left(-\frac{t}{\lambda_2}(\beta+1) \right) dt
		\\
		&=L^2\frac{1}{\lambda_2^{\beta}(\beta+1)}\exp\left(-\frac{h}{\lambda_2}(\beta+1)\right)\left(\exp\left(-\frac{\tau_1}{\lambda_2}(\beta+1) \right)-\exp\left(-\frac{\tau_2}{\lambda_2}(\beta+1) \right) \right);
		\\
		M_2&=\frac{1}{\lambda_2^2}\int_{\tau_1}^{\tau_2}\frac{1}{\lambda_2^{\beta+1}}t^2\exp\left(-\frac{h}{\lambda_2}(\beta+1)\right)\exp\left(-\frac{t}{\lambda_2}(\beta+1)\right)dt
		\\ 
		&=\frac{1}{\lambda_2^{\beta+3}}\exp\left(-\frac{h}{\lambda_2}(\beta+1)\right)\left[-\frac{\lambda_2^3}{(\beta+1)^3} \exp\left(-\frac{t}{\lambda_2}(\beta+1)\right)\left(\left(\frac{t}{\lambda_2}(\beta+1)\right)^2+\frac{2t}{\lambda_2}(\beta+1)+2\right)\right]_{\tau_1}^{\tau_2}
		\\
		&=\frac{1}{\lambda_2^{\beta}(\beta+1)^3}\exp\left(-\frac{h}{\lambda_2}(\beta+1)\right)\left( \exp\left(-\frac{\tau_1}{\lambda_2}(\beta+1)\right)\left(\left(\frac{\tau_1}{\lambda_2}(\beta+1)\right)^2+\frac{2\tau_1}{\lambda_2}(\beta+1)+2\right) \right.
		\\
		&-\left.\exp\left(-\frac{\tau_2}{\lambda_2} (\beta+1)\right)\left(\left(\frac{\tau_2}{\lambda_2}(\beta+1)\right)^2+\frac{2\tau_2}{\lambda_2}(\beta+1)\tau_2+2\right)\right);
		\\
		M_3&=\frac{1}{\lambda_2}2L\int_{\tau_1}^{\tau_2}t\frac{1}{\lambda_2^{\beta+1}}\exp\left(-\frac{h}{\lambda_2}(\beta+1)\right)\exp\left(-\frac{t}{\lambda_2}(\beta+1)\right)dt
		\\
		&=\frac{2L}{\lambda_2}\frac{1}{\lambda_2^{\beta+1}}\exp\left(-\frac{h}{\lambda_2}(\beta+1)\right)\left[-\frac{\lambda_2^2}{(\beta+1)^2}\exp\left(-\frac{t}{\lambda_2}(\beta+1)\right)\left(\frac{t}{\lambda_2}(\beta+1)+1\right)\right]_{\tau_1}^{\tau_2}
		\\
		&=\frac{2L}{\lambda_2^\beta(\beta+1)^2}\exp\left(-\frac{h}{\lambda_2}(\beta+1)\right)\left(\exp\left(-\frac{\tau_1}{\lambda_2}(\beta+1)\right)\left(\frac{\tau_1}{\lambda_2}(\beta+1)+1\right) \right.
		\\
		&-\left.\exp\left(-\frac{\tau_2}{\lambda_2}(\beta+1)\right)\left(\frac{\tau_2}{\lambda_2}(\beta+1)+1\right) \right).
	\end{align*} \par
	In conclusion, we have that:
	\begin{align*}
		J_{\tau_1, \tau_2}^\beta(a_0)&=\int _{\tau_1} ^{\tau_2} \left(\frac{\partial  \log\left(f_2(t+h|\boldsymbol{\lambda})\right)}{\partial a_0} \right)^2 f_2(t+h|\boldsymbol{\lambda})^{\beta+1} dt
		\\
		&=L^2\frac{1}{\lambda_2^{\beta}(\beta+1)}\exp\left(-\frac{h}{\lambda_2}(\beta+1)\right)\left(\exp\left(-\frac{\tau_1}{\lambda_2}(\beta+1) \right)-\exp\left(-\frac{\tau_2}{\lambda_2}(\beta+1) \right) \right)
		\\
		&+\frac{1}{\lambda_2^{\beta}(\beta+1)^3}\exp\left(-\frac{h}{\lambda_2}(\beta+1)\right)\left( \exp\left(-\frac{\tau_1}{\lambda_2}(\beta+1)\right)\left(\left(\frac{\tau_1}{\lambda_2}(\beta+1)\right)^2+\frac{2\tau_1}{\lambda_2}(\beta+1)+2\right) \right.
		\\
		&-\left.\exp\left(-\frac{\tau_2)}{\lambda_2}(\beta+1 \right)\left(\left(\frac{\tau_2}{\lambda_2}(\beta+1)\right)^2+\frac{2\tau_2}{\lambda_2}(\beta+1)+2\right)\right)
		\\
		&+\frac{2L}{\lambda_2^\beta(\beta+1)^2}\exp\left(-\frac{h}{\lambda_2}(\beta+1)\right)\left(\exp\left(-\frac{\tau_1}{\lambda_2}(\beta+1)\right)\left(\frac{\tau_1}{\lambda_2}(\beta+1)+1\right) \right.
		\\
		&-\left.\exp\left(-\frac{\tau_2}{\lambda_2}(\beta+1)\right)\left(\frac{\tau_2}{\lambda_2}(\beta+1)+1\right) \right).
	\end{align*} \par
	Finally, regarding \( J_{\tau_2}(a_0) \), we have that:
	\begin{align*}
		J^{\beta}_{\tau_2}(a_0)&=\left(\frac{\partial}{\partial a_0}\log\left(1-F_2\left(\tau_2|\boldsymbol{\lambda}\right)\right)\right)^2\left(1-F_2\left(\tau_2|\boldsymbol{\lambda}\right) \right)^{\beta+1};
		\\
		\text{But, } 1-F_2(t|\boldsymbol{\lambda})&=\exp\left(-\frac{\tau_2}{\lambda_2}-\frac{\tau_1}{\lambda_1}+\frac{\tau_1}{\lambda_2}\right);
		\log\left(1-F_2(t|\boldsymbol{\lambda})\right)=-\frac{\tau_2}{\lambda_2}-\frac{\tau_1}{\lambda_1}+\frac{\tau_1}{\lambda_2} 
	\end{align*}	
	 \par
\begin{flalign*}
    &\quad\quad \text{therefore }\quad  \frac{\partial \log\left(1-F_2(t|\boldsymbol{\lambda})\right)}{\partial a_0}=\frac{\tau_2}{\lambda_2}+\frac{\tau_1}{\lambda_1}-\frac{\tau_1}{\lambda_2}=\frac{\tau_2+h}{\lambda_2}; & \\ 
    &\quad \quad \text{and }\quad J^{\beta}_{\tau_2}(a_0)=\left(\frac{\tau_2+h}{\lambda_2}\right)^2\exp\left(-\frac{\tau_2+h}{\lambda_2}(\beta+1)\right). & 
\end{flalign*}	
 \par
	In relation to \( \xi_\beta(a_0) \), we have,
	\begin{align*}
		\xi_\beta(a_0)&=\xi_{0, \tau_1}^{\beta}(a_0)+\xi_{\tau_1, \tau_2}^{\beta}(a_0)+\xi_{\tau_2}^{\beta}(a_0).
		\\
		\noalign{\text{being:}}
		\\
		\xi_{0, \tau_1}^{\beta}(a_0)&=\int_0^{\tau_1} \left(\frac{\partial \log\left(f_1(t|\lambda_1)\right)}{\partial a_0}\right)f_1(t|\lambda_1)^{\beta+1}dt,
		\\
		\xi_{\tau_1, \tau_2}^{\beta}(a_0)&=\int_{\tau_1}^{\tau_2} \left(\frac{\partial \log\left(f_2(t+h|\boldsymbol{\lambda})\right)}{\partial a_0}\right)f_2(t+h|\boldsymbol{\lambda})^{\beta+1}dt,
		\\
		\xi_{\tau_2}^{\beta}(a_0&)= \left(\frac{\partial}{\partial a_0}\log\left(1-F_2\left(\tau_2|\boldsymbol{\lambda}\right)\right)\right)\left(1-F_2\left(\tau_2|\boldsymbol{\lambda}\right) \right)^{\beta+1}.
		\\
		\noalign{\text{Because, }}
		\xi_\beta(a_0)&=\int_0^{\tau_1} \left(\frac{\partial \log\left(f_1(t|\lambda_1)\right)}{\partial a_0}\right)f_1(t|\lambda_1)^{\beta+1}dt
		\\
		&+\int_{\tau_1}^{\tau_2} \left(\frac{\partial \log\left(f_2(t+h|\boldsymbol{\lambda})\right)}{\partial a_0}\right)f_2(t+h|\boldsymbol{\lambda})^{\beta+1}dt
		\\
		&+ \left(\frac{\partial}{\partial a_0}\log\left(1-F_2\left(\tau_2|\boldsymbol{\lambda}\right)\right)\right)\left(1-F_2\left(\tau_2|\boldsymbol{\lambda}\right) \right)^{\beta+1}
		\\
		&=\xi_{0, \tau_1}^{\beta}(a_0)+\xi_{\tau_1, \tau_2}^{\beta}(a_0)+\xi_{\tau_2}^{\beta}(a_0).
	\end{align*} 
	\par
	In relation to $\xi_{0, \tau_1}^\beta(a_0)$, we have:
	\begin{align*}
		\quad \text{Taking into } &\text{account that } \frac{\partial \log\left(f_1(t|\lambda_1\right))}{\partial a_0}=-1+\frac{t}{\lambda_1};
		\\
		\xi_{0, \tau_1}^\beta(a_0)&=\int _0 ^{\tau_1} \left(\frac{\partial  \log\left(f_1(t|\lambda_1)\right)}{\partial a_0} \right) f_1(t|\lambda_1)^{\beta+1} dt
		\\
		&=-\int _0 ^{\tau_1} f_1(t|\lambda_1)^{\beta+1} dt+\frac{1}{\lambda_1}\int _0 ^{\tau_1}t f_1(t|\lambda_1)^{\beta+1} dt
		\\
		&=-\int _0 ^{\tau_1} \frac{1}{\lambda_1^{\beta+1}} \exp\left(-\frac{t}{\lambda_1}(\beta+1)\right)dt+\frac{1}{\lambda_1}\int _0 ^{\tau_1} \frac{1}{\lambda_1^{\beta+1}} t \exp\left(-\frac{t}{\lambda_1}(\beta+1)\right) dt.
		\\
		&=-\frac{1}{\lambda_1^{ \beta}(\beta+1)} \left(1-\exp\left(-\frac{\tau_1}{\lambda_1}(\beta+1) \right) \right)+\frac{1}{\lambda_1^\beta (\beta+1)^2}\left(1-\exp\left(-\frac{\tau_1}{\lambda_1}(\beta+1) \right)\left(\frac{\tau_1}{\lambda_1}(\beta+1)+1\right)\right).
	\end{align*} \par 
	For $\xi_{\tau_1, \tau_2}^\beta(a_0)$ we have:
	\begin{align*}
		\frac{\partial \log\left(f_2(t+h|\boldsymbol{\lambda})\right)}{\partial a_0}&=L+\frac{t}{\lambda_2}
		\\
		\xi_{\tau_1, \tau_2}^\beta(a_0)&=\int _{\tau_1}^{\tau_2} \left(\frac{\partial  \log\left(f_2(t+h|\boldsymbol{\lambda})\right)}{\partial a_0} \right) f_2(t+h|\boldsymbol{\lambda})^{\beta+1} dt
		\\
		&=L\int _{\tau_1}^{\tau_2} \frac{1}{\lambda_2^{\beta+1}} \exp \left(-\frac{t+h}{\lambda_2} (\beta+1)\right)dt
		\\
		&+\frac{1}{\lambda_2}\int _{\tau_1}^{\tau_2} \frac{t}{\lambda_2^{\beta+1}} \exp \left(-\frac{t+h}{\lambda_2} (\beta+1)\right)dt
		\\
		&=L\frac{1}{\lambda_2^{\beta}(\beta+1)}\exp\left(-\frac{h}{\lambda_2}(\beta+1)\right)\left(\exp\left(-\frac{\tau_1}{\lambda_2}(\beta+1) \right)-\exp\left(-\frac{\tau_2}{\lambda_2}(\beta+1) \right) \right)
		\\
		&+\frac{1}{\lambda_2^\beta(\beta+1)^2}\exp\left(-\frac{h}{\lambda_2}(\beta+1)\right)\left(\exp\left(-\frac{\tau_1}{\lambda_2}(\beta+1)\right)\left(\frac{\tau_1}{\lambda_2}(\beta+1)+1\right) \right.
		\\
		&-\left.\exp\left(-\frac{\tau_2}{\lambda_2}(\beta+1)\right)\left(\frac{\tau_2}{\lambda_2}(\beta+1)+1\right) \right)
		\\
		&=N_1+N_2.
		\\
		N_1&=L\frac{1}{\lambda_2^{\beta}(\beta+1)}\exp\left(-\frac{h}{\lambda_2}(\beta+1)\right)\left(\exp\left(-\frac{\tau_1}{\lambda_2}(\beta+1) \right)-\exp\left(-\frac{\tau_2}{\lambda_2}(\beta+1) \right) \right);
		\\
		N_2=&=\frac{1}{\lambda_2^\beta(\beta+1)^2}\exp\left(-\frac{h}{\lambda_2}(\beta+1)\right)\left(\exp\left(-\frac{\tau_1)}{\lambda_2}(\beta+1)\right)\left(\frac{\tau_1}{\lambda_2}(\beta+1)+1\right) \right.
		\\
		&-\left.\exp\left(-\frac{\tau_2}{\lambda_2}(\beta+1)\right)\left(\frac{\tau_2}{\lambda_2}(\beta+1)+1\right) \right).
	\end{align*} \par
	Finally, we have:
	\begin{align*}
		\xi_{\tau_2}^\beta(a_0)&= \left(\frac{\partial}{\partial a_0}\log\left(1-F_2\left(\tau_2|\boldsymbol{\lambda}\right)\right)\right)\left(1-F_2\left(\tau_2|\boldsymbol{\lambda}\right) \right)^{\beta+1}
		\\
		&=\frac{\tau_2+h}{\lambda_2}\exp\left(-\frac{\tau_2+h}{\lambda_2}(\beta+1)\right).
	\end{align*} \par
	Then, $K_\beta(a_0)=J_{2\beta}(a_0)-\left(\xi_{0, \tau_1}^\beta(a_0)+\xi_{\tau_1, \tau_2}^\beta(a_0)+\xi^{\beta}_{\tau_2}(a_0)\right)^2$.
\end{proof}

\section{Proof of Corollary 3}
\begin{proof}
	Let's take \(\beta = 0\) in each expression:
	
	\begin{align*}
		&\begin{aligned}
			J^0_{0, \tau_1}(a_0) &= \exp\left(-\frac{\tau_1}{\lambda_1}\right) \left(-1 - \left(\left(\frac{\tau_1}{\lambda_1}\right)^2 + \frac{2\tau_1}{\lambda_1} + 2\right) + 2\left(\frac{\tau_1}{\lambda_1} + 1\right) \right) + 1.
			\\
			&= \exp\left(-\frac{\tau_1}{\lambda_1}\right) \left(-1 - \left(\frac{\tau_1}{\lambda_1}\right)^2 - \frac{2\tau_1}{\lambda_1} - 2 + \frac{2\tau_1}{\lambda_1} + 2 \right) + 1.
			\\
			&= \exp\left(-\frac{\tau_1}{\lambda_1}\right) \left(-1 - \left(\frac{\tau_1}{\lambda_1}\right)^2 \right) + 1.
			\\
			&= 1 - \exp\left(-\frac{\tau_1}{\lambda_1}\right) \left(1 + \left(\frac{\tau_1}{\lambda_1}\right)^2 \right).
		\end{aligned}
		\\
		&\begin{aligned}
			J^0_{\tau_1, \tau_2}(a_0)&=L^2\ \exp\left(-\frac{h}{\lambda_2}\right)\left(\exp\left(-\frac{\tau_1}{\lambda_2} \right)-\exp\left(-\frac{\tau_2}{\lambda_2} \right) \right) 
			\\
			&+\exp\left(-\frac{h}{\lambda_2}\right) \left( \exp\left(-\frac{\tau_1}{\lambda_2}\right)\left(\left(\frac{\tau_1}{\lambda_2}\right)^2+\frac{2\tau_1}{\lambda_2}+2\right) \right.
			\\
			&-\left.\exp\left(-\frac{\tau_2}{\lambda_2} \right)\left(\left(\frac{\tau_2}{\lambda_2}\right)^2+\frac{2\tau_2}{\lambda_2}+2\right)\right)
			\\
			&+2L \exp\left(-\frac{h}{\lambda_2}\right)\left(\exp\left(-\frac{\tau_1}{\lambda_2}\right)\left(\frac{\tau_1}{\lambda_2}+1\right) -\exp\left(-\frac{\tau_2}{\lambda_2}\right)\left(\frac{\tau_2}{\lambda_2}+1\right) \right).
		\end{aligned}
		\\
		&\begin{aligned}
			J^0_{\tau_2}(a_0)=\left(\frac{\tau_2+h}{\lambda_2}\right)^2\exp\left(-\frac{\tau_2+h}{\lambda_2}\right).
		\end{aligned}
	\end{align*}
	
	Similarly, for \(\xi\):
	
	\begin{align*}
		&\begin{aligned}
			\xi_{0, \tau_1}^{0}(a_0)&=- \left(1-\exp\left(-\frac{\tau_1}{\lambda_1} \right) \right)+\left(1-\exp\left(-\frac{\tau_1}{\lambda_1} \right)\left(\frac{\tau_1}{\lambda_1}+1\right)\right)
			&=-\frac{\tau_1}{\lambda_1}\exp\left(-\frac{\tau_1}{\lambda_1}\right).
		\end{aligned}
		\\
		&\begin{aligned}
			\xi_{\tau_1, \tau_2}^{0}(a_0)&=L\exp\left(-\frac{h}{\lambda_2}\right)\left(\exp\left(-\frac{\tau_1}{\lambda_2} \right)-\exp\left(-\frac{\tau_2}{\lambda_2} \right) \right) 
			\\
			&+\exp\left(-\frac{h}{\lambda_2}\right)\left(\exp\left(-\frac{\tau_1}{\lambda_2}\right)\left(\frac{\tau_1}{\lambda_2}+1\right) -\exp\left(-\frac{\tau_2}{\lambda_2}\right)\left(\frac{\tau_2}{\lambda_2}+1\right) \right)
			\\
			&=\left(-1+\frac{h}{\lambda_2}\right)\exp\left(-\frac{h}{\lambda_2}\right)\left(\exp\left(-\frac{\tau_1}{\lambda_2} \right)-\exp\left(-\frac{\tau_2}{\lambda_2} \right) \right) 
			\\
			&+\exp\left(-\frac{h}{\lambda_2}\right)\left(\exp\left(-\frac{\tau_1}{\lambda_2}\right)\left(\frac{\tau_1}{\lambda_2}+1\right) -\exp\left(-\frac{\tau_2}{\lambda_2}\right)\left(\frac{\tau_2}{\lambda_2}+1\right) \right)
			\\
			&=\left(\frac{\tau_1}{\lambda_1}-\frac{\tau_1}{\lambda_2}\right)\exp\left(-\frac{h}{\lambda_2}\right)\left(\exp\left(-\frac{\tau_1}{\lambda_2} \right)-\exp\left(-\frac{\tau_2}{\lambda_2} \right) \right) 
			\\
			&+\exp\left(-\frac{h}{\lambda_2}\right)\left(\exp\left(-\frac{\tau_1}{\lambda_2}\right)\frac{\tau_1}{\lambda_2}-
			\exp\left(-\frac{\tau_2}{\lambda_2}\right)\frac{\tau_2}{\lambda_2}\right)
			\\
			&=\frac{\tau_1}{\lambda_1}\exp\left(-\frac{\tau_1}{\lambda_1}\right)-\frac{\tau_1}{\lambda_1}\exp\left(-\frac{\tau_2+h}{\lambda_2}\right)+\frac{\tau_1}{\lambda_2}\exp\left(-\frac{\tau_2+h}{\lambda_2}\right)-\frac{\tau_2}{\lambda_2}\exp\left(-\frac{\tau_2+h}{\lambda_2}\right)
			\\
			&=\frac{\tau_1}{\lambda_1}\exp\left(-\frac{\tau_1}{\lambda_1}\right)-\left(\frac{\tau_2+h}{\lambda_2}\right)\exp\left(-\frac{\tau_2+h}{\lambda_2}\right).
		\end{aligned}
		\\
		&\begin{aligned}
			\xi_{\tau_2}^{0}(a_0) &=\left(\frac{\tau_2+h}{\lambda_2}\right)\exp\left(-\frac{\tau_2+h}{\lambda_2}\right).
		\end{aligned}
	\end{align*}
	So
	\begin{align*}
		\xi_{0}(a_0)=\xi_{0, \tau_1}^{0}(a_0)+\xi_{\tau_1, \tau_2}^{0}(a_0)+\xi_{\tau_2}^{0}(a_0)&=-\frac{\tau_1}{\lambda_1}\exp\left(-\frac{\tau_1}{\lambda_1}\right)+\frac{\tau_1}{\lambda_1}\exp\left(-\frac{\tau_1}{\lambda_1}\right)
		\\
		&-\left(\frac{\tau_2+h}{\lambda_2}\right)\exp\left(-\frac{\tau_2+h}{\lambda_2}\right)+\left(\frac{\tau_2+h}{\lambda_2}\right)\exp\left(-\frac{\tau_2+h}{\lambda_2}\right)
		\\
		&=0.
	\end{align*}
	Then, noting that 
	\[
	K_{0}(a_0) = J_{2*0}(a_0) - \xi_{0}(a_0)^2 = J_{0}(a_0),
	\]
	we obtain 
	\[
	V(\hat{a_0}^0) = J_{0}(a_0)^{-1} K_0(a_0) J_0(a_0)^{-1} = J_{0}(a_0)^{-1} J_{0}(a_0) J_0(a_0)^{-1} = J_{0}(a_0)^{-1}.
	\]
	Thus, \( J_{0}(a_0) \) corresponds to the Fisher information for \( a_0 \): 
	\begin{align*}
		J_{0}(a_0) &= \int_0^{\tau_2} \left(\frac{\partial f(t|\boldsymbol{\lambda})}{\partial a_0}\right) ^2f(t|\boldsymbol{\lambda}) dt \\ 
		&= \mathbb{E} \left[ \left( \frac{\partial}{\partial a_0} \log f(T|a_0) \right)^2 \right] = I(a_0).
	\end{align*}
\end{proof}

\section{Proof of Theorem 4}
\small
\begin{proof}
	With the considered model we have:
	\begin{align*}
		J_\beta(a_1)&=\int_0^{\tau_1} \left(\frac{\partial \log\left(f_1(t|\lambda_1)\right)}{\partial a_1}\right)^2f_1(t|\lambda_1)^{\beta+1}dt
		\\
		&+\int_{\tau_1}^{\tau_2} \left(\frac{\partial \log\left(f_2(t+h|\boldsymbol{\lambda})\right)}{\partial a_1}\right)^2f_2(t+h|\boldsymbol{\lambda})^{\beta+1}dt
		\\
		&+ \left(\frac{\partial}{\partial a_1}\log\left(1-F_2\left(\tau_2|\boldsymbol{\lambda}\right)\right)\right)^2\left(1-F_2\left(\tau_2|\boldsymbol{\lambda}\right) \right)^{\beta+1}
		\\
		&=J_{0, \tau_1}^{\beta}(a_1)+J_{\tau_1, \tau_2}^{\beta}(a_1)+J_{\tau_2}^{\beta}(a_1).
	\end{align*}
	\begin{align*}
		J_{0, \tau_1}^{\beta}(a_1)&=\int_0^{\tau_1} \left(\frac{\partial \log\left(f_1(t|\lambda_1)\right)}{\partial a_1}\right)^2f_1(t|\lambda_1)^{\beta+1}dt,
		\\
		J_{\tau_1, \tau_2}^{\beta}(a_1)&=\int_{\tau_1}^{\tau_2} \left(\frac{\partial \log\left(f_2(t+h|\boldsymbol{\lambda})\right)}{\partial a_1}\right)^2f_2(t+h|\boldsymbol{\lambda})^{\beta+1}dt,
		\\
		J_{\tau_2}^{\beta}(a_1)&= \left(\frac{\partial}{\partial a_1}\log\left(1-F_2\left(\tau_2|\boldsymbol{\lambda}\right)\right)\right)^2\left(1-F_2\left(\tau_2|\boldsymbol{\lambda}\right) \right)^{\beta+1}.
	\end{align*} \par
	$J_{0, \tau_1}^{\beta}(a_1)$:
	\begin{align*}
		\frac{\partial \log\left(f_1(t|\lambda_1) \right)}{\partial a_1}&=x_1\left(-1+\frac{t}{\lambda_1} \right)
		\\
		J_{0, \tau_1}^{\beta}(a_1)&=\int_{0}^{\tau_1}\left (\frac{\partial \log\left(f_1(t|\lambda_1) \right)}{\partial a_1}\right)^2f_1(t|\lambda_1, \lambda_2)^{1+\beta}dt
		\\
		&=x_1^2\int_0^{\tau_1}\left(-1+\frac{t}{\lambda_1}\right)^2f_1(t|\lambda_1)^{1+\beta}dt=x_1^2J_{0, \tau_1}^{\beta}(a_0).
	\end{align*} \par
	$J_{\tau_1, \tau_2}^{\beta}(a_1)$:
	\begin{align*}
		& \frac{\partial \log\left(f_2(t+h|\boldsymbol{\lambda})\right)}{\partial a_1}=\frac{\partial \log\left(f_2(t+h|\boldsymbol{\lambda})\right)}{\partial \lambda_1}\cdot \frac{\partial \lambda_1}{\partial a_1}+\frac{\partial \log\left(f_2(t+h|\boldsymbol{\lambda})\right)}{\partial \lambda_2}\cdot \frac{\partial \lambda_2}{\partial a_1} 
		\\
		& = -\frac{1}{\lambda_2}\lambda_2x_2+\frac{t}{\lambda_2^2}\lambda_2x_2-\frac{\tau_1}{\lambda_1^2}\lambda_1x_1+\frac{\tau_1}{\lambda_2^2}\lambda_2x_2=\left(-x_2-\frac{\tau_1}{\lambda_2}x_2+\frac{\tau_1}{\lambda_1}x_1 \right) +\frac{tx_2}{\lambda_2}=L^*+x_2\frac{t}{\lambda_2};
		\\
		& \Rightarrow \left(\frac{\partial  \log\left(f_2(t+h|\boldsymbol{\lambda})\right)}{\partial a_1} \right)^2=L^{*^2}+\frac{x_2^2}{\lambda_2^2}t^2+\frac{2L^*x_2}{\lambda_2}t.
	\end{align*} \par
	Therefore, we have:
	\begin{align*}
		J_{\tau_1, \tau_2}^\beta(a_1)&=\int _{\tau_1} ^{\tau_2} \left(\frac{\partial  \log\left(f_2(t+h|\boldsymbol{\lambda})\right)}{\partial a_1} \right)^2 f_2(t+h|\boldsymbol{\lambda})^{\beta+1} dt
		\\
		&=L^{*^2}\int _{\tau_1}^{\tau_2} \frac{1}{\lambda_2^{\beta+1}} \exp \left(-\frac{t+h}{\lambda_2}(\beta+1)\right)dt+\frac{x_2^2}{\lambda_2^2}\int _{\tau_1}^{\tau_2} t^2\frac{1}{\lambda_2^{\beta+1}} \exp \left(-\frac{t+h}{\lambda_2}(\beta+1)\right)dt
		\\
		&+\frac{2L^*x_2}{\lambda_2}\int _{\tau_1}^{\tau_2} \frac{t}{\lambda_2^{\beta+1}} \exp \left(-\frac{t+h}{\lambda_2}(\beta+1)\right)dt
		\\
		&=M_1^*+M_2^*+M_3^*.
	\end{align*} \par
	Where:
	\begin{align*}
		M_1^*&=\left(\frac{L^*}{L}\right)^2M_1
		\\
		M_2^*&=x_2^2M_2
		\\
		M_3^*&=\left(\frac{L^*x_2}{L}\right)M_3.
	\end{align*} \par
	Finally, we have for $J_{\tau_2}^{\beta}(a_1)$:
	\begin{align*}
		J_{\tau_2}^{\beta}(a_1)&=\left(\frac{\tau_2}{\lambda_2}x_2+\frac{\tau_1}{\lambda_1}x_1-\frac{\tau_1}{\lambda_2}x_2\right)^2\exp \left(\left(-\frac{\tau_2+h}{\lambda_2}\right)(1+\beta)\right).
	\end{align*} \par
	In relation to \( \xi_\beta(a_1) \), we have,
	\begin{align*}
		\xi_\beta(a_1)&=\xi_{0, \tau_1}^{\beta}(a_1)+\xi_{\tau_1, \tau_2}^{\beta}(a_1)+\xi_{\tau_2}^{\beta}(a_1).
		\\
		\noalign{\text{being:}}
		\\
		\xi_{0, \tau_1}^{\beta}(a_1)&=\int_0^{\tau_1} \left(\frac{\partial \log\left(f_1(t|\lambda_1)\right)}{\partial a_1}\right)f_1(t|\lambda_1)^{\beta+1}dt,
		\\
		\xi_{\tau_1, \tau_2}^{\beta}(a_1)&=\int_{\tau_1}^{\tau_2} \left(\frac{\partial \log\left(f_2(t+h|\boldsymbol{\lambda})\right)}{\partial a_1}\right)f_2(t+h|\boldsymbol{\lambda})^{\beta+1}dt,
		\\
		\xi_{\tau_2}^{\beta}(a_1)&= \left(\frac{\partial}{\partial a_1}\log\left(1-F_2\left(\tau_2|\boldsymbol{\lambda}\right)\right)\right)\left(1-F_2\left(\tau_2|\boldsymbol{\lambda}\right) \right)^{\beta+1}.
		\\
		\noalign{\text{Because, }}
		\xi_\beta(a_1)&=\int_0^{\tau_1} \left(\frac{\partial \log\left(f_1(t|\lambda_1)\right)}{\partial a_1}\right)f_1(t|\lambda_1)^{\beta+1}dt
		\\
		&+\int_{\tau_1}^{\tau_2} \left(\frac{\partial \log\left(f_2(t+h|\boldsymbol{\lambda})\right)}{\partial a10}\right)f_2(t+h|\boldsymbol{\lambda})^{\beta+1}dt
		\\
		&+ \left(\frac{\partial}{\partial a_1}\log\left(1-F_2\left(\tau_2|\boldsymbol{\lambda}\right)\right)\right)\left(1-F_2\left(\tau_2|\boldsymbol{\lambda}\right) \right)^{\beta+1}
		\\
		&=\xi_{0, \tau_1}^{\beta}(a_1)+\xi_{\tau_1, \tau_2}^{\beta}(a_1)+\xi_{\tau_2}^{\beta}(a_1).
	\end{align*} 
	In relation to $\xi_{0, \tau_1}^\beta(a_1)$, we have:
	\begin{align*}
		\frac{\partial f_1(t|\lambda_1)}{\partial a_1}&=x_1\left(-1+\frac{t}{\lambda_1}\right);
		\\
		\xi_{0, \tau_1}^\beta(a_1)&=\int _0 ^{\tau_1} \left(\frac{\partial  \log\left(f_1(t|\lambda_1)\right)}{\partial a_1} \right) f_1(t|\lambda_1)^{\beta+1} dt=x_1\xi_{0, \tau_1}^\beta(a_0).
	\end{align*} \par
	For $\xi_{\tau_1, \tau_2}^\beta(a_0)$ we have:
	\begin{align*}
		\frac{\partial \log\left(f_2(t+h|\boldsymbol{\lambda})\right)}{\partial a_0}&=L^*+\frac{x_2t}{\lambda_2}.
		\\
		\xi_{\tau_1, \tau_2}^\beta(a_0)&=\int _{\tau_1}^{\tau_2} \left(\frac{\partial  \log\left(f_2(t+h|\boldsymbol{\lambda})\right)}{\partial a_0} \right) f_2(t+h|\boldsymbol{\lambda})^{\beta+1} dt
		\\
		&=L^*\int _{\tau_1}^{\tau_2} f_2(t+h|\boldsymbol{\lambda})^{\beta+1} dt+\frac{x_2}{\lambda_1}\int _{\tau_1}^{\tau_2}t f_2(t+h|\boldsymbol{\lambda})^{\beta+1} dt
		\\
		&=N_1^*+N_2^*.
	\end{align*} \par
	With:
	\begin{align*}
		N_1^*&=\frac{L^*}{L}N_1
		\\
		N_2^*&=x_2N_2.
	\end{align*} \par
	Finally, we have:
	\begin{align*}
		\xi_{\tau_2}^\beta(a_1)&= \left(\frac{\partial}{\partial a_1}\log\left(1-F_2\left(\tau_2|\boldsymbol{\lambda}\right)\right)\right)\left(1-F_2\left(\tau_2|\boldsymbol{\lambda}\right) \right)^{\beta+1}
		\\
		&=\left(\frac{\tau_2}{\lambda_2}x_2+\frac{\tau_1}{\lambda_1}x_1-\frac{\tau_1}{\lambda_2}x_2\right)\exp\left(-\frac{\tau_2+h}{\lambda_2}(1+\beta)\right).
	\end{align*} \par
	Then, $K_\beta(a_1)=J_{2\beta}(a_1)-\left(\xi_{0, \tau_1}^\beta(a_1)+\xi_{\tau_1, \tau_2}^\beta(a_1)+\xi^{\beta}_{\tau_2}(a_1)\right)^2$.
\end{proof}

\section{Proof of Corollary 5}

\begin{proof}
	Let's take \(\beta = 0\) in each expression:
	
	\begin{align*}
		&\begin{aligned}
			J^0_{0, \tau_1}(a_1) &= x_1^2 \left( \exp\left(-\frac{\tau_1}{\lambda_1}\right) \left(-1 - \left(\left(\frac{\tau_1}{\lambda_1}\right)^2 + \frac{2\tau_1}{\lambda_1} +2 \right) + 2 \left(\frac{\tau_1}{\lambda_1} + 1 \right) \right) + 1 + 2 - 2 \right)
			\\
			&= x_1^2 \left( \exp\left(-\frac{\tau_1}{\lambda_1}\right) \left(-1 - \left(\frac{\tau_1}{\lambda_1}\right)^2 - \frac{2\tau_1}{\lambda_1} -2 + \frac{2\tau_1}{\lambda_1} +2 \right) + 1 \right)
			\\
			&=x_1^2 \left( \exp\left(-\frac{\tau_1}{\lambda_1}\right) \left(-1 - \left(\frac{\tau_1}{\lambda_1}\right)^2\right) + 1 \right)
			\\
			&=x_1^2\left(1-\exp\left(-\frac{\tau_1}{\lambda_1}\right)\left(1+\left(\frac{\tau_1}{\lambda_1}\right)^2\right)\right).
		\end{aligned}
		\\
		&\begin{aligned}
			J_{\tau_1, \tau_2}^0(a_1) &= L^{*2} \exp\left(-\frac{h}{\lambda_2}\right) \left(\exp\left(-\frac{\tau_1}{\lambda_2}\right) - \exp\left(-\frac{\tau_2}{\lambda_2}\right) \right) 
			\\
			&+x_2^2 \exp\left(-\frac{h}{\lambda_2}\right) \left( \exp\left(-\frac{\tau_1}{\lambda_2}\right) \left(\left(\frac{\tau_1}{\lambda_2}\right)^2 + \frac{2\tau_1}{\lambda_2} +2 \right) \right.
			\\
			&- \left. \exp\left(-\frac{\tau_2}{\lambda_2}\right) \left(\left(\frac{\tau_2}{\lambda_2}\right)^2 + \frac{2\tau_2}{\lambda_2} +2 \right) \right)
			\\
			&+ 2L^* x_2\exp\left(-\frac{h}{\lambda_2}\right) \left( \exp\left(-\frac{\tau_1}{\lambda_2}\right) \left(\frac{\tau_1}{\lambda_2} +1 \right) \right.
			\\
			&- \left. \exp\left(-\frac{\tau_2}{\lambda_2}\right) \left(\frac{\tau_2}{\lambda_2} +1 \right) \right).
		\end{aligned}
		\\
		&\begin{aligned}
			J_{\tau_2}(a_1) &= \left(\frac{\tau_2}{\lambda_2}x_2 + \frac{\tau_1}{\lambda_1}x_1 - \frac{\tau_1}{\lambda_2}x_2\right)^2 \exp\left(-\frac{\tau_2+h}{\lambda_2}\right).
		\end{aligned}
	\end{align*}
	
	Similarly, for \(\xi\):
	
	\begin{align*}
		&\begin{aligned}
			\xi_{0, \tau_1}^{0}(a_1) &= -x_1\left(1 - \exp\left(-\frac{\tau_1}{\lambda_1}\right) \right) +x_1\left(1 - \exp\left(-\frac{\tau_1}{\lambda_1}\right) \left(\frac{\tau_1}{\lambda_1} + 1\right)\right)=-x_1\exp\left(-\frac{\tau_1}{\lambda_1}\right)\left(\frac{\tau_1}{\lambda_1}\right).
		\end{aligned}
		\\
		&\begin{aligned}
			\xi_{\tau_1, \tau_2}^{0}(a_1) &= L^* \exp\left(-\frac{h}{\lambda_2}\right) \left(\exp\left(-\frac{\tau_1}{\lambda_2}\right) - \exp\left(-\frac{\tau_2}{\lambda_2}\right) \right) \\
			&+ x_2 \exp\left(-\frac{h}{\lambda_2}\right) \left( \exp\left(-\frac{\tau_1}{\lambda_2}\right) \left(\frac{\tau_1}{\lambda_2} +1 \right) 
			- \exp\left(-\frac{\tau_2}{\lambda_2}\right) \left(\frac{\tau_2}{\lambda_2} +1 \right) \right)
			\\ 
			&=	\left(-x_2-\frac{\tau_1}{\lambda_2}x_2+\frac{\tau_1}{\lambda_1}x_1\right)\exp\left(-\frac{h}{\lambda_2}\right) \left(\exp\left(-\frac{\tau_1}{\lambda_2}\right) - \exp\left(-\frac{\tau_2}{\lambda_2}\right) \right) \\
			&+ x_2 \exp\left(-\frac{h}{\lambda_2}\right) \left( \exp\left(-\frac{\tau_1}{\lambda_2}\right) \left(\frac{\tau_1}{\lambda_2} +1 \right) 
			- \exp\left(-\frac{\tau_2}{\lambda_2}\right) \left(\frac{\tau_2}{\lambda_2} +1 \right) \right)
			\\
			&=\left(-\frac{\tau_1}{\lambda_2}x_2+\frac{\tau_1}{\lambda_1}x_1\right)\exp\left(-\frac{h}{\lambda_2}\right) \left(\exp\left(-\frac{\tau_1}{\lambda_2}\right) - \exp\left(-\frac{\tau_2}{\lambda_2}\right) \right) \\
			&+ x_2 \exp\left(-\frac{h}{\lambda_2}\right) \left( \exp\left(-\frac{\tau_1}{\lambda_2}\right) \left(\frac{\tau_1}{\lambda_2} \right) 
			- \exp\left(-\frac{\tau_2}{\lambda_2}\right) \left(\frac{\tau_2}{\lambda_2} \right) \right)
			\\
			&=\frac{\tau_1}{\lambda_2}x_2\exp\left(-\frac{h}{\lambda_2}\right)\exp\left(-\frac{\tau_2}{\lambda_2}\right)+\frac{\tau_1}{\lambda_1}{x_1}\exp\left(-\frac{\tau_1}{\lambda_1}\right)-\frac{\tau_1}{\lambda_1}x_1\exp\left(-\frac{h}{\lambda_2}\right)\exp\left(-\frac{\tau_2}{\lambda_2}\right)
			\\
			&-x_2\exp\left(-\frac{h}{\lambda_2}\right)\exp\left(-\frac{\tau_2}{\lambda_2}\right)\frac{\tau_2}{\lambda_2}
			\\
			&=\frac{\tau_1}{\lambda_1}{x_1}\exp\left(-\frac{\tau_1}{\lambda_1}\right)-\left(\frac{\tau_2}{\lambda_2}x_2 + \frac{\tau_1}{\lambda_1}x_1 - \frac{\tau_1}{\lambda_2}x_2\right)\exp\left(-\frac{\tau_2+h}{\lambda_2}\right).
		\end{aligned}
		\\
		&\begin{aligned}
			\xi_{\tau_2}(a_1) &= \left(\frac{\tau_2}{\lambda_2}x_2 + \frac{\tau_1}{\lambda_1}x_1 - \frac{\tau_1}{\lambda_2}x_2\right) \exp\left(-\frac{\tau_2+h}{\lambda_2}\right).
		\end{aligned}
	\end{align*}
	So
	\begin{align*}
		\xi_{0}(a_1)=\xi_{0, \tau_1}^{0}(a_1)+\xi_{\tau_1, \tau_2}^{0}(a_1)+\xi_{\tau_2}^{0}(a_1)&=-\frac{\tau_1}{\lambda_1x_1}\exp\left(-\frac{\tau_1}{\lambda_1}\right)+\frac{\tau_1}{\lambda_1}x_1\exp\left(-\frac{\tau_1}{\lambda_1}\right)
		\\
		&-\left(\frac{\tau_2}{\lambda_2}x_2 + \frac{\tau_1}{\lambda_1}x_1 - \frac{\tau_1}{\lambda_2}x_2\right)\exp\left(-\frac{\tau_2+h}{\lambda_2}\right)
		\\
		&+ \left(\frac{\tau_2}{\lambda_2}x_2 + \frac{\tau_1}{\lambda_1}x_1 - \frac{\tau_1}{\lambda_2}x_2\right) \exp\left(-\frac{\tau_2+h}{\lambda_2}\right)
		\\
		&
		\\
		&=0.
	\end{align*}
	Then, noting that 
	\[
	K_{0}(a_1) = J_{2*0}(a_1) - \xi_{0}(a_1)^2 = J_{0}(a_1),
	\]
	we obtain 
	\[
	V(\hat{a_1}^0) = J_{0}(a_1)^{-1} K_0(a_1) J_0(a_1)^{-1} = J_{0}(a_1)^{-1} J_{0}(a_1) J_0(a_1)^{-1} = J_{0}(a_1)^{-1}.
	\]
	Thus, \( J_{0}(a_1) \) corresponds to the Fisher information for \( a_1 \): 
	\begin{align*}
		J_{0}(a_1) &= \int_0^{\tau_2} \left(\frac{\partial f(t|\boldsymbol{\lambda})}{\partial a_1}\right)^2 f(t|\boldsymbol{\lambda}) dt \\ 
		&= \mathbb{E} \left[ \left( \frac{\partial}{\partial a_1} \log f(T|a_1) \right)^2 \right] = I(a_1).
	\end{align*}
\end{proof}

\section{Proof of Theorem 6}

\begin{proof}
	\[
	J_\beta(a_0, a_1)=J_{0, \tau_1}^{\beta}(a_0, a_1)+J_{\tau_1, \tau_2}^{\beta}(a_0, a_1)+J_{\tau_2}^{\beta}(a_0, a_1).
	\]
	\begin{align*}
		J_\beta(a_0, a_1)&=\int_0^{\tau_1} \left(\frac{\partial \log\left(f_1(t|\lambda_1)\right)}{\partial a_1}\right)\left(\frac{\partial \log\left(f_1(t|\lambda_1)\right)}{\partial a_0}\right)f_1(t|\lambda_1)^{\beta+1}dt
		\\
		&+\int_{\tau_1}^{\tau_2} \left(\frac{\partial \log\left(f_2(t+h|\boldsymbol{\lambda})\right)}{\partial a_1}\right)\left(\frac{\partial \log\left(f_1(t|\lambda_1)\right)}{\partial a_0}\right)f_2(t+h|\boldsymbol{\lambda})^{\beta+1}dt
		\\
		&+ \left(\frac{\partial}{\partial a_0}\log\left(1-F_2\left(\tau_2|\boldsymbol{\lambda}\right)\right)\right)\left(\frac{\partial}{\partial a_1}\log\left(1-F_2\left(\tau_2|\boldsymbol{\lambda}\right)\right)\right)
		\\
		& \cdot \left(1-F_2\left(\tau_2|\boldsymbol{\lambda}\right) \right)^{\beta+1}
	\end{align*} \par
	\begin{align*}
		J_{0, \tau_1}^{\beta}(a_0, a_1)&=\int_0^{\tau_1} \left(\frac{\partial \log\left(f_1(t|\lambda_1)\right)}{\partial a_1}\right)\left(\frac{\partial \log\left(f_1(t|\lambda_1)\right)}{\partial a_0}\right)f_1(t|\lambda_1)^{\beta+1}dt
		\\
		J_{\tau_1, \tau_2}^{\beta}(a_0, a_1)&=\int_{\tau_1}^{\tau_2} \left(\frac{\partial \log\left(f_2(t+h|\boldsymbol{\lambda})\right)}{\partial a_1}\right)\left(\frac{\partial \log\left(f_1(t|\lambda_1)\right)}{\partial a_0}\right)f_2(t+h|\boldsymbol{\lambda})^{\beta+1}dt
		\\
		J_{\tau_2}^{\beta}(a_0, a_1)&= \left(\frac{\partial}{\partial a_0}\log\left(1-F_2\left(\tau_2|\boldsymbol{\lambda}\right)\right)\right)\left(\frac{\partial}{\partial a_1}\log\left(1-F_2\left(\tau_2|\boldsymbol{\lambda}\right)\right)\right)
		\\
		& \cdot \left(1-F_2\left(\tau_2|\boldsymbol{\lambda}\right) \right)^{\beta+1}
	\end{align*} \par
	\par
	Now the explicit expressions for each term will be obtained:
	\par
	For $J_{0, \tau_1}^{\beta}(a_0, a_1)$
	\begin{align*}
		J_{0, \tau_1}^\beta(a_0, a_1)&=\int_0^{\tau_1} \left(\frac{\partial \log\left(f_1(t|\lambda_1)\right)}{\partial a_1}\right)\left(\frac{\partial \log\left(f_1(t|\lambda_1)\right)}{\partial a_0}\right)f_1(t|\lambda_1)^{\beta+1}dt
		\\
		&=x_1\int_0^{\tau_1}\left(-1+\frac{t}{\lambda_1}\right)^2f_1(t|\lambda_1) dt=x_1J^{\beta}_{0, \tau_1}(a_0).
	\end{align*} \par
	And therefore, we have:
	\begin{align*}
		J_{0, \tau_1}^{\beta}(a_0, a_1)&=x_1\frac{1}{\lambda_1^\beta(\beta+1)}\left(\exp\left(-\frac{\tau_1}{\lambda_1}(\beta+1)\right)\left(-1-\frac{1}{(\beta+1)^2}\left(\left(\frac{\tau_1}{\lambda_1}(\beta+1)\right)^2+\frac{2\tau_1}{\lambda_1}(\beta+1)+2\right) \right.\right.
		\\
		&\left.\left.+\frac{2}{\beta+1}\left(\frac{\tau_1}{\lambda_1}(\beta+1)+1\right)\right)+1+\frac{2}{(\beta+1)^2}-\frac{2}{\beta+1}\right)
	\end{align*} \par
	For \( J_{\tau_1, \tau_2}(a_0) \), we have:
	\begin{align*}
		J_{\tau_1, \tau_2}^\beta(a_0)&=\int _{\tau_1} ^{\tau_2} \left(\frac{\partial  \log\left(f_2(t+h|\boldsymbol{\lambda})\right)}{\partial a_0} \right) \left(\frac{\partial  \log\left(f_2(t+h|\boldsymbol{\lambda})\right)}{\partial a_1} \right) f_2(t+h|\boldsymbol{\lambda})^{\beta+1} dt
		\\
		&=\int_{\tau_1}^{\tau_2} \left(L^*+x_2\frac{t}{\lambda_2}\right)\left(L+\frac{t}{\lambda_2}\right)
		\\
		&=L\cdot L^* \cdot \int _{\tau_1}^{\tau_2} \frac{1}{\lambda_2^{\beta+1}} \exp \left(-\frac{t+h}{\lambda_2} \left( \beta+1\right)\right)dt+\frac{1}{\lambda_2^2}\int _{\tau_1}^{\tau_2} x_2t^2\frac{1}{\lambda_2^{\beta+1}} \exp \left(-\frac{t+h}{\lambda_2} \left( \beta+1\right)\right)dt
		\\
		&+\frac{Lx_2+L^*}{\lambda_2}\int _{\tau_1}^{\tau_2} \frac{t}{\lambda_2^{\beta+1}} \exp \left(-\frac{t+h}{\lambda_2} \left(\beta+1\right)\right)dt
		\\
		&=L\cdot L^* M_1+x_2M_2+\left(Lx_2+L^*\right)M_3
		\\
		M_1&=L^*\cdot L \frac{1}{\lambda_2^{\beta}(\beta+1)}\exp\left(-\frac{h}{\lambda_2}(\beta+1)\right)\left(\exp\left(-\frac{\tau_1}{\lambda_2}(\beta+1) \right)-\exp\left(-\frac{\tau_2}{\lambda_2}(\beta+1) \right) \right);
		\\
		M_2&=x_2\frac{1}{\lambda_2^{\beta}(\beta+1)^3}\exp\left(-\frac{h}{\lambda_2}(\beta+1)\right)\left( \exp\left(-\frac{\tau_1}{\lambda_2}(\beta+1)\right)\left(\left(\frac{\tau_1}{\lambda_2}(\beta+1)\right)^2+\frac{2\tau_1}{\lambda_2}(\beta+1)+2\right) \right.
		\\
		&-\left.\exp\left(-\frac{\tau_2}{\lambda_2} (\beta+1)\right)\left(\left(\frac{\tau_2}{\lambda_2}(\beta+1)\right)^2+\frac{2\tau_2}{\lambda_2}(\beta+1)\tau_2+2\right)\right);
		\\
		M_3&=\frac{1}{\lambda_2^\beta(\beta+1)^2}\exp\left(-\frac{h}{\lambda_2}(\beta+1)\right)\left(\exp\left(-\frac{\tau_1}{\lambda_2}(\beta+1)\right)\left(\frac{\tau_1}{\lambda_2}(\beta+1)+1\right) \right.
		\\
		&-\left.\exp\left(-\frac{\tau_2}{\lambda_2}(\beta+1)\right)\left(\frac{\tau_2}{\lambda_2}(\beta+1)+1\right) \right).
	\end{align*} \par
	In conclusion, we have that:
	\begin{align*}
		J_{\tau_1, \tau_2}^\beta(a_0, a_1)&=\int _{\tau_1} ^{\tau_2} \left(\frac{\partial  \log\left(f_2(t+h|\boldsymbol{\lambda})\right)}{\partial a_0} \right)^2 f_2(t+h|\boldsymbol{\lambda})^{\beta+1} dt
		\\
		&=L^*\cdot L \frac{1}{\lambda_2^{\beta}(\beta+1)}\exp\left(-\frac{h}{\lambda_2}(\beta+1)\right)\left(\exp\left(-\frac{\tau_1}{\lambda_2}(\beta+1) \right)-\exp\left(-\frac{\tau_2}{\lambda_2}(\beta+1) \right) \right)
		\\
		&+\frac{x_2}{\lambda_2^{\beta}(\beta+1)^3}\exp\left(-\frac{h}{\lambda_2}(\beta+1)\right)\left( \exp\left(-\frac{\tau_1}{\lambda_2}(\beta+1)\right)\left(\left(\frac{\tau_1}{\lambda_2}(\beta+1)\right)^2+\frac{2\tau_1}{\lambda_2}(\beta+1)+2\right) \right.
		\\
		&-\left.\exp\left(-\frac{\tau_2)}{\lambda_2}(\beta+1 \right)\left(\left(\frac{\tau_2}{\lambda_2}(\beta+1)\right)^2+\frac{2\tau_2}{\lambda_2}(\beta+1)+2\right)\right)
		\\
		&+\frac{L^*+x_2L}{\lambda_2^\beta(\beta+1)^2}\exp\left(-\frac{h}{\lambda_2}(\beta+1)\right)\left(\exp\left(-\frac{\tau_1}{\lambda_2}(\beta+1)\right)\left(\frac{\tau_1}{\lambda_2}(\beta+1)+1\right) \right.
		\\
		&-\left.\exp\left(-\frac{\tau_2}{\lambda_2}(\beta+1)\right)\left(\frac{\tau_2}{\lambda_2}(\beta+1)+1\right) \right).
	\end{align*} \par
	Finally, regarding \( J_{\tau_2}(a_0, a_1) \), we have that:
	\begin{align*}
		J^{\beta}_{\tau_2}(a_0, a_1)&=\left(\frac{\partial}{\partial a_0}\log\left(1-F_2\left(\tau_2|\boldsymbol{\lambda}\right)\right)\right)\left(\frac{\partial}{\partial a_1}\log\left(1-F_2\left(\tau_2|\boldsymbol{\lambda}\right)\right)\right)\left(1-F_2\left(\tau_2|\boldsymbol{\lambda}\right) \right)^{\beta+1};
		\\
		&=\left(\frac{\tau_2}{\lambda_2}x_2+\frac{\tau_1}{\lambda_1}x_1-\frac{\tau_1}{\lambda_2}x_2\right)\left(\frac{\tau_2+h}{\lambda_2}\right)\exp \left(-\left(-\frac{\tau_2+h}{\lambda_2}\right)(1+\beta)\right).
	\end{align*} \par
	In relation to \( \xi_\beta(a_0, a_1) \), we have,
	\begin{align*}
		\xi_\beta(a_0, a_1)=\left(\xi_\beta(a_0), \xi_\beta(a_1)\right).
	\end{align*}
	Then:
	\[
	K_{\beta}(a_0, a_1)=J_{2\beta}(a_0, a_1)-\xi_{\beta}(a_0, a_1)^\top \xi_{\beta}(a_0, a_1)
	\]
\end{proof}

\section{Proof of Corollary 7}
\begin{proof}
	Let's take \(\beta = 0\) in each expression:
	
	\begin{align*}
		&\begin{aligned}
			J^0_{0, \tau_1}(a_0, a_1)  &=x_1\left(1-\exp\left(-\frac{\tau_1}{\lambda_1}\right)\left(1+\left(\frac{\tau_1}{\lambda_1}\right)^2\right)\right).
		\end{aligned}
		\\
		&\begin{aligned}
			J_{\tau_1, \tau_2}^0(a_0a_1) &= L^*\cdot L \exp\left(-\frac{h}{\lambda_2}\right) \left(\exp\left(-\frac{\tau_1}{\lambda_2}\right) - \exp\left(-\frac{\tau_2}{\lambda_2}\right) \right) 
			\\
			&+x_2 \exp\left(-\frac{h}{\lambda_2}\right) \left( \exp\left(-\frac{\tau_1}{\lambda_2}\right) \left(\left(\frac{\tau_1}{\lambda_2}\right)^2 + \frac{2\tau_1}{\lambda_2} +2 \right) \right.
			\\
			&- \left. \exp\left(-\frac{\tau_2}{\lambda_2}\right) \left(\left(\frac{\tau_2}{\lambda_2}\right)^2 + \frac{2\tau_2}{\lambda_2} +2 \right) \right)
			\\
			&+ \left(L^*+x_2L\right)\exp\left(-\frac{h}{\lambda_2}\right) \left( \exp\left(-\frac{\tau_1}{\lambda_2}\right) \left(\frac{\tau_1}{\lambda_2} +1 \right) \right.
			\\
			&- \left. \exp\left(-\frac{\tau_2}{\lambda_2}\right) \left(\frac{\tau_2}{\lambda_2} +1 \right) \right)
		\end{aligned}
		\\
		&\begin{aligned}
			J_{\tau_2}(a_0, a_1) &= \left(\frac{\tau_2}{\lambda_2}x_2 + \frac{\tau_1}{\lambda_1}x_1 - \frac{\tau_1}{\lambda_2}x_2\right)\left(\frac{\tau_2+h}{\lambda_2}\right) \exp\left(-\frac{\tau_2+h}{\lambda_2}\right).
		\end{aligned}
	\end{align*}
	
	Similarly, for \(\xi\):
	
	\[
	\xi_0(a_0, a_1)=(0, 0).
	\]
	So
	Then, noting that 
	\[
	K_{0}(a_0, a_1) = J_{2*0}(a_0, a_1) - \xi_{0}(a_0, a_1)^\top \xi_{0}(a_0, a_1) = J_{0}(a_0, a_1),
	\]
	we obtain 
	\[
	J_{0}(a_0, a_1)^{-1} K_0(a_0, a_1)J_0(a_0, a_1)^{-1} = J_{0}(a_0, a_1)^{-1} J_{0}(a_0, a_1) J_0(a_0, a_1)^{-1} = J_{0}(a_0, a_1)^{-1}.
	\]
	Thus, \( J_{0}(a_0, a_1) \) corresponds to the Fisher information for \(\left(a_0, a_1\right) \): 
	\begin{align*}
		J_{0}(a_0, a_1) &= \int_0^{\tau_2}\left(\frac{\partial f(t|\boldsymbol{\lambda})}{\partial a_0}\right) \left(\frac{\partial f(t|\boldsymbol{\lambda})}{\partial a_1}\right) f(t|\boldsymbol{\lambda}) dt \\ 
		&= \mathbb{E} \left[ \left( \frac{\partial}{\partial a_0} \log f(T|a_1, a_0) \right)\left( \frac{\partial}{\partial a_1} \log f(T|a_1, a_0) \right)\right] = I(a_0, a_1).
	\end{align*}
\end{proof}

\end{document}